\documentclass[aps,pra,twocolumn,showpacs,preprintnumbers,amsmath,amssymb,superscriptaddress,longbibliography]{revtex4-1}
\usepackage[english]{babel}
\usepackage{graphicx}
\usepackage{dcolumn}
\usepackage{bm}

\begin{document}

\preprint{APS/123-QED}

\title{Light mediated non-Gaussian atomic ensemble entanglement}

\author{Olov Pettersson}
\affiliation{Department of Theoretical Physics, Royal Institute of Technology, Stockholm, Sweden}
\affiliation{New York University Shanghai, 1555 Century Ave, Pudong, Shanghai 200122, China}  

\author{Tim Byrnes}%
\affiliation{New York University Shanghai, 1555 Century Ave, Pudong, Shanghai 200122, China}
\affiliation{State Key Laboratory of Precision Spectroscopy, School of Physical and Material Sciences,
East China Normal University, Shanghai 200062, China}
\affiliation{NYU-ECNU Institute of Physics at NYU Shanghai, 3663 Zhongshan Road North, Shanghai 200062, China}
\affiliation{National Institute of Informatics, 2-1-2 Hitotsubashi, Chiyoda-ku, Tokyo 101-8430, Japan}
\affiliation{Department of Physics, New York University, New York, NY 10003, USA}

\date{\today}

\begin{abstract}
We analyze a similar scheme for producing light-mediated entanglement between atomic ensembles, as first realized by Julsgaard, Kozhekin and Polzik [Nature {\bf 413}, 400 (2001)]. In the standard approach to modeling the scheme, a Holstein-Primakoff approximation is made, where the atomic ensembles are treated as bosonic modes, and is only valid for short interaction times.  In this paper, we solve the time evolution without this approximation, which extends the region of validity of the interaction time.  For short entangling times, we find this produces a state with similar characteristics as a two-mode squeezed state, in agreement with standard predictions.  For long entangling times, the state evolves into a non-Gaussian form, and the two-mode squeezed state characteristics start to diminish.  This is attributed to more exotic types of entangled states being generated. We characterize the states by examining the Fock state probability distributions, Husimi $Q$ distributions, and non-local entanglement between the ensembles.  We compare and connect several quantities obtained using the Holstein-Primakoff approach and our exact time evolution methods. 
\end{abstract}

\pacs{03.75.Pp, 03.75.Be, 03.67.Bg, 03.67.Mn}
\maketitle

\section{Introduction}
\label{sec:intro}

Quantum phenomena are typically associated with the microscopic world where delicate quantum states are thought to be impossible to generate for all but the smallest number of particles. In the last few decades, macroscopic systems have been created in which large number of particles behave quantum mechanically. Some remarkable examples which have been achieved experimentally are cantilevers being in superposition of oscillation modes \cite{OConnell2010}, non-classical state generation and teleportation of macroscopic atomic ensembles 
\cite{Julsgaard2001a,Riedel2010,Krauter2013,pezze16}. On the other hand it is well-known that macroscopic quantum systems often suffer from decoherence exponentially with particle number \cite{RevModPhys.76.1267,RevModPhys.75.715} and the quantum mechanical properties of the state can only be sustained for very short times. This apparent inconsistency is resolved by understanding that decoherence is a state-dependent process.  Some states such as Schrodinger cat states are highly susceptible to decoherence, while coherent states are relatively robust.  By using quantum states which are less affected by decoherence, this gives the possibility of realistically realizing macroscopic states that can be used for various applications such as quantum metrology and quantum information \cite{Byrnes2013,Byrnes2015,Byrnes2012,filip13,Ilo-Okeke2014,1367-2630-16-7-073038,tichy16,PhysRevA.90.062336}.  

\begin{figure}
    \includegraphics[width=\linewidth]{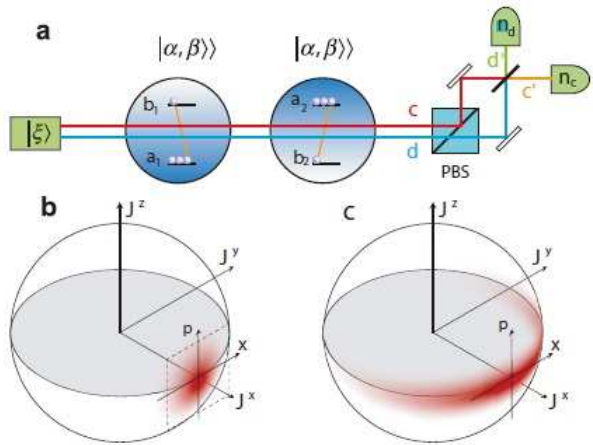}
    \caption{(a) Schematic of the entanglement process. A two-mode coherent light pulse $|\xi\rangle$ is sent through two atomic ensembles of cold atoms initially in coherent spin states polarized in opposing $ J^x $ directions, entangling the two ensembles and light. The light modes are then interfered via a beamsplitter after passing through a polarizing beam splitter (PBS), with measured photon numbers $n_c$ and $n_d$ in each mode. (b) (c) The Bloch sphere representation for a state (b) well-approximated and (c) beyond the Holstein-Primakoff (HP) approximation for spin states.   The HP approximation maps a region around a polarized spin on the Bloch sphere into a flat phase space. For large deviations from the polarized ground state, the mapping become increasingly inaccurate.}
    \label{fig:schematics}
\end{figure}

Currently, the types of states used in large ensemble states fall under two basic categories, continuous variable (CV) \cite{Furusawa23101998,Braunstein2005} or discrete quantum states. CV has been extensively used to model entanglement and teleportation using degrees of freedom analogous to quadratures of light, and has been extremely successful in demonstrating entanglement generation and teleportation between multiple ensembles \cite{Julsgaard2001a,Krauter2013}.   Here the transversal spin operators, normally ($J^z$ and $J^y$) of a $J^x$ polarized spin ensemble are treated as canonical position and momentum quadratures of the quantum harmonic oscillator. Deviations from the polarized state on the Bloch sphere, are mapped onto position and momentum variables according to the Holstein-Primakoff (HP) approximation $ x = J^y/\sqrt{2N} $ and $ p = J^z/\sqrt{2N} $, where $ N $ is the particle number.  The approximation as a single quantum harmonic oscillator mode is accurate as long as the quantum states is close to the original $J^x$ polarization (Fig. \ref{fig:schematics}(b)).  However, for states that deviate from near the $J^x$ polarization the approximation breaks down and the full spin degrees of freedom must be considered (Fig. \ref{fig:schematics}(c)). We call such a regime where the HP approximation can no longer be used the "beyond-CV" regime, as a correct description must involve keeping track of spin, rather than mode operators.  

In experiments such as Ref. \cite{Julsgaard2001a}, the type of entangled state is a two-mode squeezed state under the  HP approximation. In this experiment, described schematically in Fig. \ref{fig:schematics}(a),  the entanglement is generated by letting two polarization modes of a laser interact with two atomic ensembles via the ac Stark shift \cite{Kuzmich2000,Duan2001,Duan2000,N.J.CerfG.Leuchs}. A pair of measurements is then made on the light after interfering the modes.  While it is well-established that the procedure creates two-mode squeezed states in the ensembles, given that the ensembles are not genuine quantum harmonic oscillator modes, a question remains of what happens in the light-matter interaction is strong enough such that the HP approximation breaks down.  This question is particularly relevant as recently several experiments have reported the generation of non-Gaussian spin ensemble states \cite{gerving2012,strobel2014,barontini15,mcconnell15,bohnet16,pezze16}. For our purposes we define non-Gaussian states as those with quasiprobability distributions (i.e. the Wigner or Husimi $Q$-distribution) that cannot be described by a Gaussian form. An important question in this context is: what are the characteristics of the states that are generated by the scheme of Fig. \ref{fig:schematics}(a) in the beyond-CV regime?

In this paper, we investigate the quantum states created by a similar protocol to that introduced in Ref. \cite{Julsgaard2001a} but without using the HP approximation. The procedure is described in Fig. \ref{fig:schematics}(a), where initially atom-light-atom entanglement is created due to an ac Stark shift interaction.  The HP approximation restricts the region of validity to only short interaction times between the atoms and the light.  Using our exact time evolution techniques, this allows us to consider long interaction times which can produce states that requires one to use the full spin formalism.  The light is then measured, which produces atom-atom entanglement.  We analyze the resultant state by examiing the probability distributions, Husimi $Q$-distributions, and entanglement.  

This paper is structured as follows.  In Sec. \ref{sec:discrete} we first review the standard approach using the HP approximation, and derive a procedure that maps between quantum states on the atomic ensembles and the approximated mode representation.  In Sec. \ref{sec:theory} we construct a theory of measurement induced entanglement for the scheme shown in Fig. \ref{fig:schematics}(a).  Our theory of entanglement generation between two atomic ensembles works in the regime where the HP approximation no longer is valid. In Sec. \ref{sec:approximation}, we obtain approximate expressions by analytic methods to find the relation between the two-mode squeezed state and the qubit entangled state. In Sec. \ref{sec:q-functions} we plot the probability density and Husimi $Q$-distributions \cite{citeulike:7123199} of the entangled state. In doing so, we reveal the properties of the entangled state in the beyond-CV regime, where it differs significantly from a two-mode squeezed state. In Sec. \ref{sec:entanglement} we analyze the atomic entangled state in terms of the entanglement generation between the ensembles.  We finally give our conclusions in Sec. \ref{sec:discussion}.

\section{Continuous variable approach to light mediated atomic ensemble entanglement}
\label{sec:discrete} 

We now review the standard approach of describing atomic ensemble entanglement based on the technique as described in Refs. \cite{Kuzmich2000,Duan2000,Julsgaard2001a,Gross2012,Hammerer2010}. The approach is based on making a HP approximation to the spin variables of the ensembles, and results in two-mode squeezing interaction.  After describing the general framework and results of this approach, we describe what this approximation means in terms of the atomic states such that the approach can be compared to our methods introduced in later sections.

\subsection{Standard continuous variables approach}
\label{sec:standard}

In this section we review the standard theoretical approach of atomic ensemble entanglement based on the HP approximation. In this approach, the discrete collective spin operators are mapped to the continuous canonical position and momentum operators of the harmonic oscillator with unit mass,
\begin{equation}\label{eq:cvmapping}
x = \frac{J^y}{\sqrt{2N}}, \qquad p =  \frac{J^z}{\sqrt{2N}},
\end{equation}
which obey the usual commutation relation $[x,p] = i$. This approximation maps a part of the Bloch sphere around the polarized spin vector in which deviations are discrete into the flat phase space of continuous variables, seen in Fig. \ref{fig:schematics}(b). Intuitively it is clear that this approximation only is valid when the angular deviation is small, as for larger deviations the mapping become increasingly unsatisfactory, as seen (exaggerated) in Fig. \ref{fig:schematics}(c). In this framework, an entangled state is modeled as a two-mode squeezed state in which the variance of the total position and relative momentum tend to zero with increased squeezing. The atom-light interaction in the scheme described in Fig. \ref{fig:schematics} creates a time evolution in which the states of light and matter are entangled, and a measurement of one reveals information about the other.

In this scheme, an interaction Hamiltonian of the form 
\begin{align}
H = gS^{z}(J_{1}^{z} + J_{2}^{z}) \nonumber
\end{align}
is used, where $S^z$ is the Stokes operator describing a light source and $J^{z}$ are the atomic Schwinger boson operators describing the atomic systems. Using the same mapping as in Eq. (\ref{eq:cvmapping}) for both the atomic and photonic operators, we define the canonical continuous variable operators  as  $x_{\text{at}}$, $p_{\text{at}} $ and $x_{\text{ph}}$, $p_{\text{ph}}$ respectively. As the laser pulse passes through the atomic samples, the photonic operators evolve according to Heisenberg's equation of motion, 
\begin{equation}
\begin{aligned}
&\hbar \frac{dx_{\text{ph}}}{dt} = ig[p_{\text{ph}}(p_{\text{at},1} + p_{\text{at},2}),x_{\text{ph}}] = g(p_{\text{at},1} + p_{\text{at},2}), \\
 &\hbar  \frac{dp_{\text{ph}}}{dt} = ig[p_{\text{ph}}(p_{\text{at},1} + p_{\text{at},2}),p_{\text{ph}}] = 0.
\end{aligned}
\end{equation}
Assuming the operators remain stationary in time for the duration of the experiment, we get the output modes by simple integration as,
\begin{align}
x_{\text{ph}}(t) & \approx x_{\text{ph}}(0) + \frac{gt}{\hbar} \big(p_{\text{at},1}(0) + p_{\text{at},2} (0) \big), \label{eq:photonevolved1} \\
p_{\text{ph}} (t) & \approx p_{\text{ph}} (0). \label{eq:photonevolved2}
\end{align}
Similarly, the atomic operators evolve as \cite{Kuzmich2000, Duan2000}
\begin{align}
x_{\text{at},1}(t) & \approx x_{\text{at},1}(0) + \frac{gt}{\hbar} p_{\text{ph}} (0), \label{eq:atomicevolved1} \\
x_{\text{at},2}(t) &  \approx x_{\text{at},2} (0) - \frac{gt}{\hbar} p_{\text{ph}} (0) , \label{eq:atomicevolved2} \\
 p_{\text{at},1} (t) & \approx p_{\text{at},1} (0) ,\label{eq:atomicevolved3}\\
 p_{\text{at},2} (t)& \approx p_{\text{at},2} (0) .
\label{eq:atomicevolved4}
\end{align}
After the evolution a measurement of $x_{\text{ph}}$ is performed, it is measured and collapses to a constant.  According to (\ref{eq:photonevolved1}) and the fact that 
\begin{align}
p_{\text{at},1} (t) +  p_{\text{at},2} (t) = p_{\text{at},1} (0) + p_{\text{at},2} (0)
\end{align}
from (\ref{eq:atomicevolved3}) and (\ref{eq:atomicevolved4}) this means that  the quantity  $p_{\text{at},1}(t) + p_{\text{at},2}(t)$ also collapses to a constant, provided $ x_{\text{ph}}(0) $ is small and the dimensionless interaction parameter $gt/\hbar$  is large.  This means that the variance of $p_{\text{at},1}(t) + p_{\text{at},2}(t)$ is small, which results in correlations between these observables.  Meanwhile, from (\ref{eq:atomicevolved1}) and (\ref{eq:atomicevolved2}) we have 
\begin{align}
x_{\text{at},1} (t) +  x_{\text{at},2} (t) = x_{\text{at},1} (0) + x_{\text{at},2} (0)
\end{align}
which means that the correlations between $ p_{\text{at},1} (t) $ and $ p_{\text{at},2} (t) $  can be produced without changing the correlations between $ x_{\text{at},1} (t) $ and $ x_{\text{at},2} (t) $.  

The procedure is then repeated again but in a different basis in order to generate the correlations between 
$ x_{\text{at},1}$ and $ x_{\text{at},2}$.  
The two atomic systems are rotated around the the $J^x$-axis, which transforms the canonical operators for the first system as $x_{\text{at},1} \to -p_{\text{at},1}$  and $p_{\text{at},1} \to x_{\text{at},1}$, while the second system is rotated in the opposite direction, $x_{\text{at},2} \to p_{\text{at},2}$, and $p_{\text{at},2} \to -x_{\text{at},2}$. A second measurement of the light field, $x_{\text{ph}}$ under the same conditions as before fixes the quantity $x_{\text{at},2} - x_{\text{at},1}$ to a constant, and gives a small variance. Similiarly to Eqs. (\ref{eq:atomicevolved1})-(\ref{eq:atomicevolved4}) the evolved quantities $p_{\text{at},1} + p_{\text{at},2} $ and $x_{\text{at},2} - x_{\text{at},1}$ remain conserved through the procedure.

\subsection{Mapping between atomic spin states and the HP approximated mode representation}
\label{sec:mapping}

In the continuous variable approximation, one of the total spin operators $ J^x $ is set to a constant and the other two operators $ J^y,J^z $ are treated as approximate position and momentum variables. Typically the effect of the atom and light interaction is derived in the Heisenberg picture where it is relatively simple to derive the entanglement.  In this section we describe what this approximation means in the Schrodinger picture, which will be the framework that will be used in our analysis in following sections.  We deduce the mapping between the original spin and HP approximated bosonic mode Fock states.  

First let us write the total spin operators in terms of bosonic modes, which are conventionally defined as
\begin{align}
J^{x} & = a^{\dagger}b  + b^{\dagger}a \nonumber \\
J^{y} & = -ia^{\dagger}b + ib^{\dagger}a \nonumber \\
J^{z} & = a^{\dagger}a - b^{\dagger}b  .
\end{align}
These obey commutation relations $ [J^l,J^m] = 2 i \epsilon_{lmn} J^n $ where $ \epsilon_{lmn} $ is the Levi-Civita antisymmetric tensor. Here, $a^{\dagger}$ and $b^{\dagger}$ denote the creation operators for two hyperfine levels of the atoms in each of the ensembles labeled by $ j$. These operators obey bosonic commutation relations $[a,a^{\dagger}] = [b,b^{\dagger}] = 1 $. To explicitly observe the HP approximation in terms of bosonic modes, we make a SU(2) transformation 
\begin{equation}
\begin{aligned}
a &= \frac{1}{\sqrt{2}}(a_x-ib_x),\\
b &= \frac{1}{\sqrt{2}}(a_x+ib_x),
\end{aligned}
\end{equation}
which transforms the spin operators as
\begin{align}
J^{x} & = a_x^{\dagger}a_x  - b_x^{\dagger} b_x \nonumber  \\
J^{y} & = a_x^{\dagger}b_x + b_x^{\dagger}a_x \nonumber \\
J^{z} & = -ia_x^{\dagger}b_x + ib_x^{\dagger}a_x . 
\end{align}
The HP approximation requires that we are in the vicinity of a fully $ J^x $ polarized state, which is
\begin{align}
|J^x=-N \rangle = \frac{1}{\sqrt{N!}}  (b_x^{\dagger})^N |0 \rangle . 
\end{align}
Due to the macroscopic population of the $ b_x $ state, we may take this mode to be a constant $ b_x \sim \sqrt{N} $.  We thus have
\begin{align}
J^{x} & \approx  a_x^{\dagger} a_x  -N \nonumber \\
J^{y} & \approx  \sqrt{N}(a_x + a_x^{\dagger}) = \sqrt{2N} x \nonumber \\
J^{z} & \approx  \sqrt{N}( -ia_x + ia_x^{\dagger}) = -\sqrt{2N} p .
\end{align}
where the position and momentum operators are $ x = ( a_x + a_x^{\dagger})/\sqrt{2} $ and $ p = ( a_x - a_x^{\dagger})/\sqrt{2}i $ \cite{Braunstein2005}.  We thus see that the HP approximation amounts to setting macroscopically occupied mode to a constant, and treating the remaining mode quantum mechanically.  

Given a particular quantum state, we may then transform between the original spin representation and the HP approximated state in the following way.  Consider an arbitrary state of the spin ensemble
\begin{align}
| \psi \rangle = \sum_{k} \psi_k | k \rangle
\end{align}
where 
\begin{align}
| k \rangle = \frac{(a^{\dagger})^{k}(b^{\dagger})^{(N-k)}}{\sqrt{k!(N-k)!}}|0\rangle
\end{align}
are the Fock states for the spins.  According to the above procedure, we must set the $ b_x $ mode to a constant, hence we must first make a change of basis to the $ J^x $ Fock states defined as 
\begin{align}
| k_x \rangle = \frac{(a_x^{\dagger})^{k}(b_x^{\dagger})^{(N-k)}}{\sqrt{k!(N-k)!}}|0\rangle.  
\label{kxdef}
\end{align}
The state is then written
\begin{align}
| \psi \rangle =  \sum_{k_x} \psi'_{k_x} | k_x \rangle
\end{align}
where $ \psi'_{k_x} = \sum_{k} \langle k_x |  k \rangle \psi_k $. The matrix elements $\langle k_x |k\rangle $ are calculated according to Refs. \cite{tinkham2003group,thompson2008angular}.  Setting $ b_x =\sqrt{N} $ in (\ref{kxdef}) results in an unnormalized state, 
hence we make the association
\begin{align}
| k_x  \rangle\text{     (spins)}  \leftrightarrow | k_{\text{ph}} \rangle \equiv \frac{1}{\sqrt{k!}} (a_x^\dagger)^k | 0 \rangle \text{     (mode)} .
\label{mapping}
\end{align}
We may thus say that the spin Fock states in the $ J^x $ basis are the bosonic mode Fock states in the HP approximation.

\section{Theory of light mediated atomic ensemble entanglement}
\label{sec:theory}

In this section we introduce our approach of deriving the ensemble-ensemble entanglement in the Schrodinger picture.  As shown in Fig. \ref{fig:schematics}(a), linearly polarized coherent light illuminates two ensembles, each in a coherent spin state.  A beam splitter then interferes the light, after which the photons are measured. As described in Sec. \ref{sec:standard}, the entanglement procedure makes use of two successive measurements of the light field, with the atomic ensembles rotated in between in order to infer correlation in both the $ J^z $ and $ J^y $  directions. As the two procedures are identical up to a basis rotation, we restrict our analysis to one of the measurements, which will prove to have rich dynamics as the system is evolved beyond the HP approximation.  

The initial state of light is  polarized in the $ S^x $-direction, which is a superposition of left and right circularly polarized light.  The initial quantum state of the light can thus be written
\begin{equation}
| \psi_{\text{light}} (t=0) \rangle = | \xi \rangle = e^{-\frac{|\xi|^{2}}{2}}\exp\Big[\frac{\xi}{\sqrt{2}}(c^{\dagger}+d^{\dagger})\Big]|0\rangle
\end{equation}
where $\xi$ is the amplitude of the light, and $ c^\dagger ,d^\dagger  $ are the creation operators for left and right hand circularly polarized light.  
The Stokes operators are defined as
\begin{align}
S^{x} & = c^{\dagger}d  + d^{\dagger}c \nonumber \\
S^{y} & = -ic^{\dagger}d + id^{\dagger}c \nonumber \\
S^{z} & = c^{\dagger}c - d^{\dagger}d  .
\end{align}
The initial state of light has the expectation value $ \langle  S^x \rangle = |\xi|^2 $ but is zero for $ \langle  S^{y,z} \rangle = 0 $. 

For the atomic ensembles, the initial state is 
\begin{equation}
| \psi_{\text{atoms}} (t=0) \rangle = 
|\frac{\pi}{2}, 0 \rangle\rangle_1 |\frac{\pi}{2},\pi \rangle\rangle_2
\label{atominitial}
\end{equation}
where we have defined the  coherent spin states as
\begin{equation}
\label{eq:cssactual}
|\theta,\phi \rangle \rangle_j = \prod_{n=1}^{N_j} \left( \cos \left(\tfrac{\theta}{2} \right)    | a_j \rangle_n +  
\sin \left( \tfrac{\theta}{2} \right) e^{i \phi} | b_j \rangle_n \right) .
\end{equation}
Here $ | a \rangle_n, | b \rangle_n $ denote the two hyperfine states of the $ n $th atom in the ensemble, and $ \theta $ and $ \phi $ are two arbitrary angles on the Bloch sphere, $ 0 < \theta < \pi $, $ 0 < \phi < 2 \pi $.
The index $j = 1,2 $ labels the two ensembles, and $ N_j $ is the number of atoms in each ensemble. In our calculations, it will be convenient to work with the bosonic formulation of coherent spin states, which is valid as long as the wavefunction in the ensemble is symmetric under particle interchange.  This will be always true in our case, as the Hamiltonian used to evolve the system and the initial states are symmetric. The spin coherent state in the bosonic formulation is written
\begin{equation}
\label{eq:css}
|\theta,\phi  \rangle \rangle_j = \frac{1}{\sqrt{N!}}(\cos \left(\tfrac{\theta}{2} \right)    a^{\dagger}_j + \sin \left( \tfrac{\theta}{2} \right) e^{i \phi }  b^{\dagger}_j)^{N_j} |0\rangle .
\end{equation}
where the bosonic operators $ a_j, b_j $ are defined as in Sec. \ref{sec:mapping}, with the additional $ j =1,2 $ labels for each ensemble. We note that in Ref. \cite{Julsgaard2001a} in fact the ground state of the underlying atoms is a $ F=4 $ state, and our case above would strictly speaking correspond to $ F = 1/2 $.  While these may appear different, another way to view each ensemble is that it is a macroscopic spin with total spin $ J = N $, where all the underlying spins are symmetric under particle interchange. In this picture the constituent particles making up the macroscopic spin become irrelevant to the dynamics as long we make observations in the total spin variable.  In this sense our results should be also valid for any $ F $ making up the ensemble.   

With the use of the spin operators we model the interaction between the two atomic ensembles as a non-linear quantum non-destructive (QND) Hamiltonian of the form  \cite{Kuzmich2007,Takahashi1999,Takeuchi2005}
\begin{equation}
\label{eq:hamiltonian}
H = g S^{z}(J_{1}^{z} + J_{2}^{z}),
\end{equation}
where $g$ is the interaction parameter resulting from the ac Stark shift coupling. In terms of physical parameters, this is given by $ g = \frac{\sigma \gamma \alpha_v}{A(I+\frac{1}{2})\Delta}$ where $\sigma$ is the resonant absorption cross section for an unpolarized photon on an unpolarized atom, $A$ is the cross section area of the light beam, $\gamma$ the spontaneous emission rate from the upper atomic level, $\alpha_v$ the vector polarizability and $I$ the value of the nuclear spin \cite{Kuzmich2000}. The interaction is considered to be a QND if $[H,S^z]= 0$, which is satisfied in our case. 

We evolve the system in time by applying the unitary time evolution operator $U = \exp(-it H/\hbar)$ which yields, with the dimensionless entanglement time $\tau = gt/\hbar $,
\begin{align}
|\Psi(\tau)\rangle & = \exp(-it H/\hbar) | \psi_{\text{atoms}} (t=0) \rangle | \psi_{\text{light}} (t=0) \rangle  \nonumber \\
& = \frac{1}{\sqrt{2^{(N_1+N_2)}}}   \sum_{k_{1},k_{2}}\sqrt{C_{N_1}^{k_{1}}C_{N_2}^{k_{2}}}(-1)^{k_2}\nonumber \\ 
& \times e^{-i\tau(2k_{1} + 2k_{2} -(N_1+N_2))(c^\dagger c - d^\dagger d) }|k_{1},k_{2}\rangle |\xi\rangle
\end{align}
where we have expanded the atomic states in the $J^z_i $ eigenbasis. The light component of this can be evaluated by noting that a number operator on a state causes a coherent state to pick up a phase $e^{i \theta c^\dagger c } e^{\xi c^\dagger} | 0 \rangle  = \exp[ \xi e^{i\theta} c^\dagger ] | 0 \rangle  $. This shifts the phase of the optical state with a value dependent on the $ J^z_j $ eigenstate.  
Now the light and atoms become entangled, giving
\begin{align}
 |\Psi(\tau)\rangle & = \frac{e^{-\frac{|\xi|^{2}}{2}}}{\sqrt{2^{(N_1+N_2)}}}
\sum_{k_{1},k_{2}}\sqrt{C_{N_1}^{k_{1}}C_{N_2}^{k_{2}}}(-1)^{k_2} \nonumber \\
& \times \exp\Big[\frac{\xi}{\sqrt{2}}  \big(e^{-i\tau(2k_{1} + 2k_{2} - (N_1+N_2))}c^{\dagger} \nonumber \\
&  + e^{i\tau(2k_{1} + 2k_{2} - (N_1+N_2))}d^{\dagger}\big)\Big]|k_{1},k_{2}\rangle|0\rangle  .
\end{align}
The phases picked up by the interaction are now interfered using a using a 50:50 beam splitter after the light pulse has passed through the two atomic ensembles. The beam splitter transforms the photonic operators as 
\begin{align}
c^{\dagger} & = \frac{1}{\sqrt{2}}(c'^{\dagger} + id'^{\dagger})  \nonumber \\
d^{\dagger} & = -\frac{1}{\sqrt{2}}(ic'^{\dagger} +d'^{\dagger}),
\end{align}
which yields
\begin{align}
& |\Psi(\tau)\rangle = \frac{e^{-\frac{|\xi|^{2}}{2}}}{\sqrt{2^{(N_1+N_2)}}}\sum_{k_{1},k_{2}}\sqrt{C_{N_1}^{k_{1}}C_{N_2}^{k_{2}}}(-1)^{k_2} |k_{1},k_{2}\rangle \nonumber  \\
& \times 
\exp\Big[-\xi e^{\frac{i\pi}{4}} i\sin(x + \frac{\pi}{4})c'^{\dagger} + \xi e^{\frac{i\pi}{4}}i\cos(x + \frac{\pi}{4})d'^{\dagger}\Big]
|0\rangle
\end{align}
where $x = \tau(2k_{1} + 2k_{2} - (N_1+N_2))$. The last step is to project the above state on the photonic number states $|n_c,n_d\rangle$, 
with measurement outcomes $n_c$ and $n_d$ respectively. We note here that depending on whether a photon resolving measurement is made, $ n_c $ and $ n_d $ may or may not be explicitly known.  Regardless of whether this is known, we shall see that entanglement will be produced between the atomic ensembles, on a shot-to-shot level.  

The state after projecting on the photonic number states is
\begin{multline}
\label{eq:aes}
|\Psi(\tau)\rangle =  \frac{1}{\sqrt{{\cal N}}}  \frac{1}{\sqrt{ 2^{(N_1+N_2)} }} \sum_{k_{1},k_{2}}
\sqrt{C_{N_1}^{k_{1}}C_{N_2}^{k_{2}}}(-1)^{k_2}\\
\times A_{n_{c} n_{d}}(k_{1},k_{2})|k_{1},k_{2}\rangle
\end{multline}
where
\begin{equation} 
\begin{aligned}
\label{eq:a}
A_{n_{c} n_{d}}(k_{1},k_{2}) &= \frac{e^{-\frac{|\xi|^2}{2}}\xi^{n_{c} + n_{d}}}{\sqrt{n_{c}!}\sqrt{n_{d}!}}\\
\times & \sin^{n_{c}}(\tau(2k_{1} + 2k_{2} - (N_1+N_2)) + \frac{\pi}{4})\\ 
\times & \cos^{n_{d}}(\tau(2k_{1} + 2k_{2} - (N_1+N_2))  + \frac{\pi}{4}),
\end{aligned}
\end{equation}
and
\begin{align}
{\cal N} & =  \frac{1}{ 2^{(N_1+N_2)} }  \sum_{k_{1},k_{2}} C_{N_1}^{k_{1}} C_{N_2}^{k_{2}} A_{n_{c} n_{d}}^2  (k_{1},k_{2})
\end{align}
is a normalization factor as a measurement of the light component was made, and irrelevant global phase factors were dropped. Eq. (\ref{eq:aes}) is the core result of this paper, the latter sections are devoted to examining its properties. For zero interaction time $ \tau = 0 $, $ A_{n_c,n_d}(k_1,k_2)$ is constant in $k_1$ and $k_2$ and the state of the atoms are unaffected by the measurement. Evolving in $\tau$ gives entanglement 
between the two ensembles due to the correlations in $ A_{n_c,n_d}(k_1,k_2) $.  In the next section we analyze this function to see what kind of correlations are present due to the atom-light interaction.

\section{Analytic approximation of probability densities}
\label{sec:approximation}

As Eq. (\ref{eq:aes}) is the wavefunction after the interaction with the light, in principle it is possible to derive all physical quantities based on this.  However, due to the complicated nature of the coefficient $ A_{n_{c} n_{d}}(k_{1},k_{2})  $ it is not entirely obvious what kind of entangled state this is and hence it is beneficial to make some analytical approximations.   For small light-ensemble interaction times, we expect the atomic entangled state to behave like a two-mode squeezed state, as the HP approach should be a good approximation in this regime. First, the binomial coefficient can with good approximation be treated as Gaussian when $N\gg1$,
\begin{align}
\label{eq:binomialapprox}
\frac{C^{k}_{N}}{2^N} \approx \sqrt{\frac{2}{N\pi}}\exp[-\frac{1}{2N}(2k - N)^2 ] .  
\end{align}
Secondly, using Stirling approximation we may write for $ N_p \gg 1 $ the combination of the binomial and trigonometric functions as a Gaussian function \cite{Ilo-okeke2015},
\begin{multline}
\label{eq:ebube}
C^{n_c}_{N_p}\cos^{2n_c}(x+\frac{\pi}{4})\sin^{2N_p-2n_c}(x+\frac{\pi}{4}) \\
 \approx \frac{2}{| \cos(2x)|} \exp\Big[-\frac{2 N_p}{\cos^2(2x) } \Big(\frac{2n_c-N_p}{2N_p} - \frac{1}{2}|\sin (2x)| \Big)^2 \Big]
\end{multline}
with $x=2\tau(k_1+k_2-N)$. Here, the measured photon numbers in each mode are $n_c$ and $n_d$ and thus the total photon number is $n_c + n_d = N_p$. The maximum of the Gaussian occurs at
\begin{align}
\sin[ 4\tau(k_1 + k_2 -N)]  =\frac{2n_c-N_p}{N_p},
\label{maxcondition}
\end{align}
which has in general multiple solutions for $ k_1, k_2 $, due to the oscillatory nature of the sine function.  As the beam splitter in Fig. \ref{fig:schematics}(a) is a 50:50 beam splitter, on average we expect that the number of photons in each detector is equal $ n_c \approx n_d \approx N_p/2 $.  Restricting our analysis to small times $ \tau \sim 1/N $, we may linearize the trigonometric functions around $ x \ll 1 $, and obtain the simplified expression for the probability distribution,
\begin{multline}
p(k_1,k_2) \propto \exp \left(- 2N \left[ \left(\frac{2k_{1}-N}{2N} \right)^2 + \left(\frac{2k_{2}-N}{2N} \right)^2\right] \right)\\
\times \exp \Big( -8N_p \tau^2 (k_{1} + k_{2} -N )^2  \Big).
\label{approxprob}
\end{multline}
The first factor in this expression states that atom Fock states occur with averages $\langle k_{1,2} \rangle = N/2$ and with standard deviation $\sqrt{N}/2$.  This is the same distribution as the original spin coherent states (\ref{atominitial}) that the atoms are initially prepared in.  The second factor gives a correlation between the two atomic modes which results in entanglement. The strength of the correlation is dependent on the square of the dimensionless entanglement time and the number of photons in the interaction. 

\begin{figure}
    \includegraphics[width=0.48\textwidth]{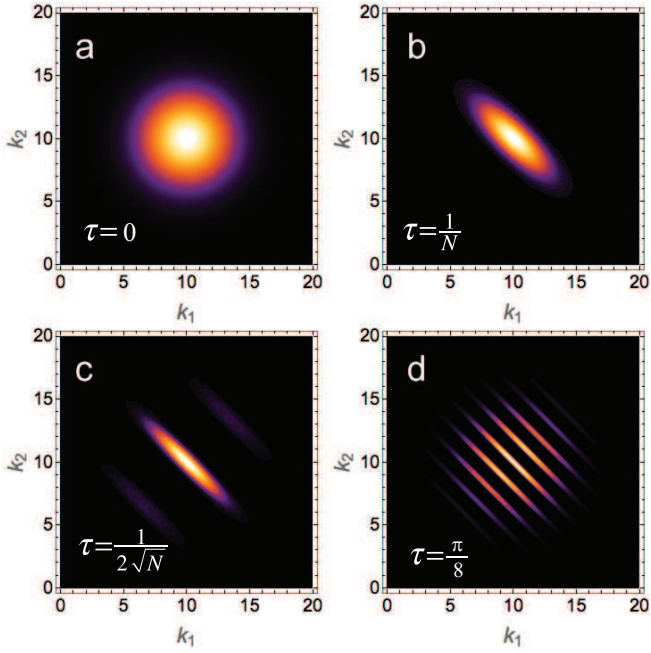}
    \caption{Probability densities of the light mediated atomic entangled state, using Eq. (\ref{eq:aes}) and the approximations in Eq. (\ref{eq:binomialapprox}) and (\ref{eq:ebube}), for various interaction times $\tau$.  Parameters used are (a) $\tau=0$, (b) $\tau=1/N$, (c) $\tau = 1/2\sqrt{N}$, (d) $\tau = \pi/8$. Parameters used are $N=20$, $n_c=n_d=10$.}
    \label{fig:aes-prob}
\end{figure}

\begin{figure}
    \includegraphics[width=0.48\textwidth]{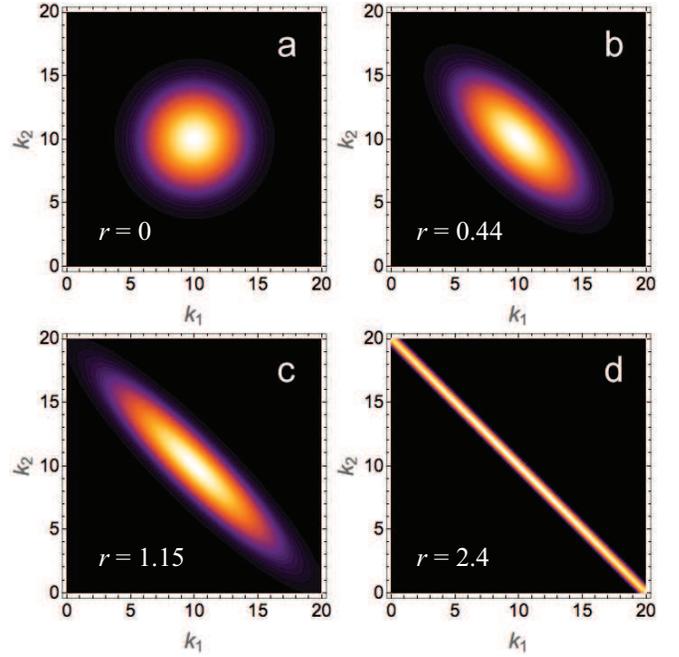}
    \caption{Probability densities for the mapped two-mode squeezed state,  Eq. (\ref{tmss-final}).  Squeezing parameters used are (a) $r=0 $, (b) $r=0.44$, (c) $r=1.15$, 
(d) $r=2.40$ and $ N = 20 $.}
    \label{fig:aes-probmapped}
\end{figure}

To confirm our analytical prediction, we numerically plot the probability distribution of the wavefunction in Eq. (\ref{eq:aes}) for different entanglement times in Fig. \ref{fig:aes-prob}. When $\tau=0$, there is no correlation between the two modes and the probability distribution is a Gaussian distribution centered around $k_1 = k_2 = N/2$. Evolving the interaction to times of order $\tau \sim 1/N$, the state develops correlations between Fock states of the ensembles, in agreement with (\ref{approxprob}).  This state should be consistent with a two-mode squeezed state, according to the arguments of Sec. \ref{sec:standard}.  To verify this, let us write the wavefunction of the two-mode squeezed state 
\begin{align}
|\text{TMS (mode)} \rangle = \sqrt{1-\tanh^2(r)} \sum_{n=0}^\infty \tanh^n(r) |n \rangle |n \rangle 
\label{tmsphoton}
\end{align}
where the $ |n \rangle $ are bosonic mode Fock states. To compare the probability distributions of the states, we employ the mapping procedure as discussed in Sec. \ref{sec:mapping}, where the mode Fock states are mapped back onto the spin Fock states in the $ J^x $ basis.  We thus expect that the two-mode squeezed state has a state in terms of the spin states according to 
\begin{align}
|& \text{TMS (spin)}\rangle  = \frac{1}{\sqrt{{\cal N_{\text{TMS}} }}} \sum_{k_x=0}^N \tanh^{k_x}(r) |k_x \rangle |k_x \rangle  \nonumber \\
& = \frac{1}{\sqrt{{\cal N_{\text{TMS}} }}} \sum_{k_x,k_1,k_2=0}^N \tanh^{k_x}(r) \langle k_1 | k_x \rangle \langle k_2 | k_x \rangle | k_1 \rangle | k_2 \rangle 
\label{tmsspin}
\end{align}
where we have restricted the sum to $ N $ as now these are now Fock states for the spins, and $ {\cal N_{\text{TMS}} } $ is a suitable normalization constant. In the limit $ N \gg 1 $, it is possible to evaluate the sum over $ k $ analytically (see Appendix), and we obtain
\begin{multline}
\label{tmss-final}
p(k_1,k_2) \propto \\
\exp \left( -2N e^{-2r}  \left[ \left(\frac{2k_{1} - N}{2N} \right)^2 + \left(\frac{2k_{2} -N}{2N} \right)^2 \right]  \right)\\\
\times \exp \left(   -\frac{2 \sinh(2r) }{N}  (k_{1} + k_{2} -N)^2 \right)
\end{multline}
We see an immediate similarity of the mapped two-mode squeezed state probability distribution to (\ref{approxprob}).  As with the case of the atomic entangled state, there is a Gaussian envelope from the first exponential factor with $\langle k_{1,2} \rangle = N/2$ and standard deviation $\sqrt{N}/2$. The second exponential induces a correlation between the two modes $k_1$ and $k_2$ with squeezing parameter $r$. Increasing the squeezing parameter $r$ causes the coefficient of the second Gaussian $\sinh(2r)$ to increase exponentially, narrowing the distribution in the direction $k_1 - k_2 -N=0$ while extending it in the opposite direction. Atomic ensembles have a fixed atom number which restricts the range of $k_{1,2}$, while bosonic modes do not have this restriction. This means that the mode Fock state occupation probability extends to infinity for the HP approximated case. Inspecting the two probability density functions, we can infer a relation between the entanglement time $\tau$ and the squeezing parameter,
\begin{equation}
4 N_p N \tau^2 \leftrightarrow \sinh (2r)  .
\label{squeezingmapping}
\end{equation}
For small squeezing $ r \ll 1 $, we may approximate $ \sinh (2r)  \approx 2r $, hence we obtain $ 2N_p N \tau^2 \leftrightarrow r $ in this regime.  In Fig. \ref{fig:aes-prob} and Fig. \ref{fig:aes-probmapped} we compare the probabilities of the atomic states (\ref{eq:aes}) after the interaction with the light with the mapped two-mode squeezed state.  We see in general good agreement between the two distributions, with the expected correlations developing with squeezing parameter $ r $.


\begin{figure}
    \includegraphics[width=0.48\textwidth]{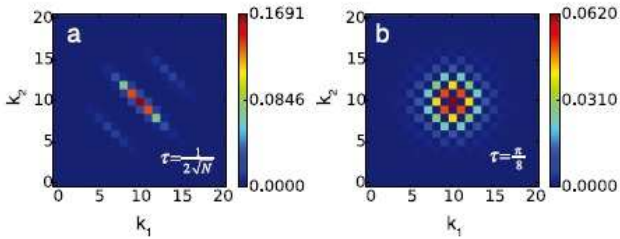}
    \caption{Exact probability density for the atomic entangled state, using Eq. (\ref{eq:aes}), for entanglement times (a) $\tau=\frac{1}{2\sqrt{N}}$, (b) $\tau=\frac{\pi}{8}$. The latter entanglement time result in a Schrodinger cat state. Parameters used are  $N=20$, $n_c = n_d = 10$}
    \label{fig:catstate}
\end{figure}

Further increasing the entanglement time $ \tau $, the Fock state correlations become stronger, which the distribution narrowing as would be expected of (\ref{approxprob}).  For interaction times of order $ \tau \sim 1 $, multiple probability distribution peaks appear indicating non-Gaussian behavior (Fig. \ref{fig:aes-prob}(c)(d)), due to the periodicity of the trigonometric functions in
(\ref{maxcondition}).  The boundary between the linear regime where only one and many correlation peaks occurs is the time 
$\tau \sim 1/\sqrt{N}$, where weak secondary peaks start to appear in Fig. \ref{fig:aes-prob}(c). The corresponding exact probability distribution is shown in Fig. \ref{fig:catstate}(a), showing good agreement to Fig. \ref{fig:aes-prob}(c)  After these times the two probability distributions diverge, and the atomic entangled state entangles $k$-states other than $k_1 + k_2 = N$, which is apparent from Fig. \ref{fig:aes-prob}(c)(d). These correlations beyond the primary peak emerge when there are more than one solution for the maximum of (\ref{eq:ebube}). The multiple peaks signal the breakdown of the HP approximation, where the correlations cannot be treated linearly. 

One difference between the approximated distribution and the atomic entangled state is that probability distributions of the former tend to broaden as the number correlations become stronger, while the latter distributions stay within a fixed envelope.  This occurs because in the infinitely squeezed limit the mapped two-mode squeezed state (\ref{tmsspin}) is 
\begin{align}
| \text{TMS (spin } r \rightarrow \infty \text{)}   \rangle & = \frac{1}{\sqrt{N+1}} \sum_{k_x=0}^N | k_x \rangle | k_x \rangle \nonumber \\
& = \frac{1}{\sqrt{N+1}} \sum_{k=0}^N | k \rangle | k \rangle  . 
\label{eprstate}
\end{align}
The probability distribution is thus a perfectly Fock number correlated state with a uniform distribution from $ k=0 $ to $ N$.  Such a state never truly occurs in the atomic ensemble wavefunction, due to the development of multiple correlation peaks within a finite envelope probability.  The closest distribution to this would be that shown in Fig. \ref{fig:aes-prob}(b), where there is one correlation peak and the times are still within the HP approximated regime.  From (\ref{squeezingmapping}) we can estimate the maximum squeezing that is possible
\begin{align}
\sinh (2 r_{\max}) \approx \frac{2N_p}{N} .
\label{maxsqueezing}
\end{align}
This suggests that it is advantageous to use very bright light in comparison to the ensemble particle number in order to achieve a high squeezing.  We note that (\ref{maxsqueezing}) should be viewed as a theoretical upper bound to the squeezing and in practice there will be other factors such as decoherence which will reduce the maximum possible squeezing further.

At even longer times, the lines become increasingly dense until $\tau=\frac{\pi}{8}$, where the probability becomes a checkerboard pattern, where every second $k$-state has a non-zero probability, as seen in Fig. \ref{fig:catstate}. Inserting this time, $\tau = \frac{\pi}{8}$ into (\ref{eq:aes}), while assuming $n_c = n_d$, a sum of trigonometric functions, with $m=k_1+k_2-N+1$ is obtained,
\begin{equation}
\begin{aligned}
|\Psi\rangle &\propto \frac{1}{2^N}\sum_{k_1,k_2} \sqrt{C_N^{k_1}C_N^{k_2}} \sin^{n_c}\big(\frac{m\pi}{4}\big)\cos^{n_d}\big(\frac{m\pi}{4}\big)|k_1,k_2\rangle \\
&= \frac{1}{2^{n_c}}\frac{1}{2^N}\sum_{k_1,k_2} \sqrt{C_N^{k_1}C_N^{k_2}} \cos^{n_c}\big(\frac{m\pi}{2}\big)|k_1,k_2\rangle. \\
\end{aligned}
\end{equation}
Depending on the parity of $n_c$, this evaluates in two different ways. For even $n_c$, $ \cos^{n_c}\big(\frac{m\pi}{2}\big) = \frac{1}{2}(1 + (-1)^{m})$, while for odd $n_c$, $ \cos^{n_c}\big(\frac{m\pi}{2}\big) = \frac{1}{2}(i^m +(-i)^m)$. We then have, for odd $n_c$,
\begin{equation}
\begin{aligned}
|\Psi\rangle &\propto \frac{1}{2^{n_c}}\frac{1}{2^N}\sum_{k_1,k_2} \sqrt{C_N^{k_1}C_N^{k_2}}\frac{1}{2}(i^m +(-i)^m) |k_1,k_2\rangle \\
&= \frac{i^{-N+1}}{2}\frac{1}{2^N}\sum_{k_1,k_2} \sqrt{C_N^{k_1}C_N^{k_2}}i^{k_1} (-i)^{k_2}|k_1,k_2\rangle \\
&+ \frac{i^{-N+1}}{2}\frac{1}{2^N}\sum_{k_1,k_2} \sqrt{C_N^{k_1}C_N^{k_2}}(-i)^{k_1} i^{k_2}|k_1,k_2\rangle.
\end{aligned}
\end{equation}
Dropping overall constant factors and using Eq. (\ref{eq:cssactual}), we can identify the sums as coherent spin states at distinct positions on the Bloch sphere. The odd state then becomes 
\begin{equation}
\begin{aligned}
|\Psi\rangle &\propto  |\dfrac{\pi}{2} ,-\dfrac{\pi}{2} \rangle\rangle |\dfrac{\pi}{2} ,-\dfrac{\pi}{2} \rangle\rangle 
+|\dfrac{\pi}{2} ,\dfrac{\pi}{2} \rangle\rangle |\dfrac{\pi}{2} ,\dfrac{\pi}{2} \rangle\rangle, 
\end{aligned}
\label{eq:catstate-odd}
\end{equation}
while the case for even $n_c$ results in a similar state,
\begin{equation}
|\Psi\rangle \propto|\dfrac{\pi}{2} ,0 \rangle\rangle |\dfrac{\pi}{2} ,0 \rangle\rangle     + |\dfrac{\pi}{2} ,\pi \rangle\rangle |\dfrac{\pi}{2} ,\pi \rangle\rangle  .  
\label{eq:catstate-even}
\end{equation}
Now we see that for this particular time, assuming the measured photon numbers in each mode are equal, we obtain Bell states consisting of states at opposite sides of the Bloch sphere. Since the superposition of states involves spin coherent states at opposite sides of the Bloch sphere,  we call the states (\ref{eq:catstate-odd}) and (\ref{eq:catstate-even}) "Schrodinger cat-Bell states". However, the axis in which the states are polarized depends on the parity of the $n_c$. For odd values, the state end up in $\pm J^y$, while for even numbers, the direction is switched to $\pm J^x$. A similar form of entangled states was obtained for a pure $ J^z_1 J^z_2 $ interaction in Ref. \cite{Byrnes2013}.  At a time $\tau=\frac{\pi}{4}$ the atomic states become disentangled again, and the states are unchanged from the initial state $ | \psi_{\text{atoms}} (t=0) \rangle $.

\section{Husimi $Q$-distributions}
\label{sec:q-functions}

In the previous section we saw that the light mediated entanglement produced a state with a probability distribution consistent with a two-mode squeezed state for small interaction times, and a more complex correlation structure for longer times.  In order to understand the nature of the correlations we analyze the atomic entangled state in terms of coherent spin state Husimi $Q$-distributions, defined as
\begin{align}
Q & ( \theta_1, \phi_1, \theta_2, \phi_2 ) = \frac{1}{\pi^2} | \langle \langle \theta_1, \phi_1 |  \langle \langle \theta_2, \phi_2 | \Psi \rangle |^2   .
\label{qfuncall}
\end{align}
As the state that we wish to analyze has two ensembles, the $Q$-function has four degrees of freedom $ \theta_1, \phi_1, \theta_2, \phi_2 $, and cannot be visualized directly.  For this reason we consider the marginal and joint $Q$-functions, where one of the ensembles is traced over in the former case, and the conditional distribution is examined in the latter.  The resulting $Q$-distributions then become a function of two variables which can be visualized.

\subsection{Marginal $Q$-functions}

\begin{figure}
    \includegraphics[width=0.48\textwidth]{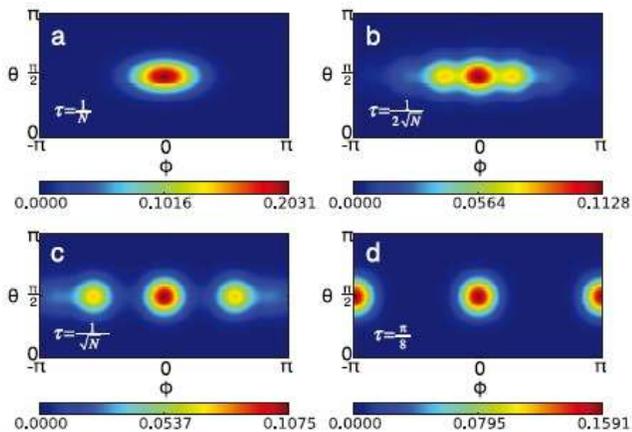}
    \caption{Marginal $Q$-distributions for the entangled state, Eq. (\ref{eq:q-marginal-css}), at various entanglement times $\tau$. (a) $\tau=1/N$, (b) $\tau = 1/2\sqrt{N}$, (c) $\tau = 1/\sqrt{N}$, (d) $\tau = \pi/8$. In all plots, the atomic particle number $N=20$, and measured photon numbers $n_c$ = $n_d$ = 10 }
    \label{fig:q-css-marginal}
\end{figure}

The marginal $Q$-function is obtained by tracing out one of the ensembles of the atomic entangled state
\begin{align}
\rho_1 = \text{Tr}_2 ( | \Psi \rangle \langle \Psi | ) = \sum_{k_2} | \langle k_2 | \Psi \rangle |^2 ,
\label{partialtrace}
\end{align}
and finding the overlap with coherent spin states as defined in (\ref{eq:css}).  We obtain
\begin{equation}
\begin{aligned}
\label{eq:q-marginal-css}
Q& (\theta_1,\phi_1) = \frac{1}{\pi}\langle \langle \theta_1, \phi_1 | \rho_1 |\theta_1, \phi_1  \rangle \rangle \\
&=\frac{1}{\pi 4^N }\sum_{k_2}C_N^{k_2} \\
& \times \Big|\sum_{k_1} C_N^{k_1} 
e^{i\phi(N-k_1)} \cos^{k_1} (\tfrac{\theta}{2}) \sin^{N-k_1} (\tfrac{\theta}{2})
A(k_1,k_2) \Big|^2.
\end{aligned}
\end{equation}
In Fig. \ref{fig:q-css-marginal} we see the behavior of the marginal $Q$-function for different interaction times. The $Q$-function starts as a symmetric Gaussian at $ \tau = 0 $, as the initial state is simply a coherent spin state and is independent of the second ensemble \cite{Byrnes2012}
\begin{align}
Q (\theta_1,\phi_1) \approx \frac{1}{\pi} \exp \left(-\frac{N(\theta_1 - \pi/2)^2}{4}-\frac{N\phi_1^2}{4} \right).
\end{align}
For short times $\tau = 1/N$, the $Q$-function starts to broaden in the $ \phi $ direction, due to the $ J^z_1+J^z_2 $ Hamiltonian which rotates the spins around the $ z $ axis of the Bloch sphere.  The partial trace has the effect of collapsing all the various terms in the entangled state, creating a broadening effect, in a similar way to Ref. \cite{Byrnes2013}.  Up until the characteristic time $\tau \lesssim 1/\sqrt{N}$, the coherent states are all distributed around the original point in the Bloch sphere to a high degree, due to the fact that $A(k_1,k_2)$ in (\ref{eq:a}) varies little for different $k$ in the sum. Further evolving in $\tau$ creates Gaussians centered at different positions around the equator of the Bloch sphere, in analogy to that seen in Ref. \cite{Byrnes2013}.  At $\tau = \pi/8$ there are two Gaussian distributions at $ \phi= 0, \pi $ and $ \theta = \pi/2 $, which we attribute to the Schrodinger cat entangled state (\ref{eq:catstate-even}). Finally the $Q$-functions are periodic in time $\tau =\pi/4$, as would be expected from (\ref{eq:css}).  

Let us now compare the $Q$-functions to a genuine two-mode squeezed state to see the degree of agreement. The marginal $Q$-function (in the space of bosonic modes) of the two-mode squeezed state is
\begin{align}
Q(\alpha) = \frac{1}{\pi} \langle \alpha | \rho_1^{\text{TMS}} | \alpha \rangle
\end{align}
where the coherent state is 
\begin{align}
|\alpha\rangle = e^{\frac{-|\alpha|^2}{2}}\sum_n \frac{\alpha^n}{\sqrt{n!}}|n\rangle
\end{align}
and
\begin{align}
\rho_1^{\text{TMS}} & = \text{Tr}_1 ( | \text{TMS (mode)} \rangle  \langle \text{TMS (mode)} | ) \nonumber \\
& = (1- \tanh^2 (r)) \sum_{n=0}^\infty \tanh^{2n} (r) |n \rangle \langle n | .
\label{photondens}
\end{align}
This gives the $Q$-function
\begin{align}
Q(\alpha) =\frac{(1- \tanh^2 (r))}{\pi} e^{ -|\alpha|^2 (1- \tanh^2(r)) },
\end{align}
which is a Gaussian function of radius $ \sim 1/(1- \tanh^2(r)) $.  We compare this by applying the spin-mode mapping from section \ref{sec:mapping} to calculate the $Q$-function (for bosonic modes) of the mapped atomic entangled state. The first step of the mapping is to transform the state into the $ J^x $ basis, giving
\begin{equation}
\begin{aligned}
|\Psi(\tau) \rangle &= \frac{1}{2^N} \sum_{k_1  k_2 k^x_1} 
\sqrt{C_N^{k_1} C_N^{k_2} } A(k_1,k_2) (-1)^{k_2} \langle k^x_1|k_1\rangle  |k^x_1\rangle | k_2\rangle .
\end{aligned}
\end{equation}
We then interpret the $ | k_x \rangle $ states to be the bosonic mode Fock state, which allows us to evaluate the $Q$-function to be
\begin{align}
Q & (\alpha) = \frac{1}{\pi 4^N } e^{-|\alpha|^2}  \nonumber \\
& \times \sum_{k_2} \Big|\sum_{k^x_1 k_1} 
\sqrt{C_N^{k_1} C_N^{k_2} } A(k_1,k_2) \langle k^x_1 |k_1 \rangle 
\frac{\alpha^{k^x_1}}{\sqrt{k^x_1!}} \Big|^2.
\label{eq:q-marginal-optical}
\end{align}
This distribution is defined in the phase space of bosonic modes as a function of the complex parameter $\alpha$. 

\begin{figure}
    \includegraphics[width=0.48\textwidth]{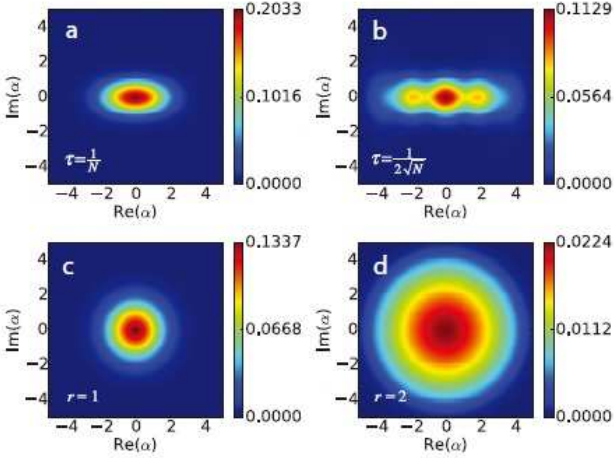}
    \caption{(a)(b) Marginal $Q$-distributions (for bosonic modes) for the transformed entangled state, Eq. (\ref{eq:q-marginal-optical}), at various entanglement times (a) $\tau=1/N$, (b) $\tau = 1/2\sqrt{N}$ for particle number $N=20$. (c)(d)   Comparison with $Q$-functions for a genuine two-mode squeezed state for the marginal Q-distributions, Eq. (\ref{photondens}), with parameters (c) $r=1.0$  (d) $r=2.0$.    }
    \label{fig:q-optical-marginal}
\end{figure}

The comparison between the two distributions is shown in Fig. \ref{fig:q-optical-marginal}. We see that for the genuine two-mode squeezed state, the marginal $Q$-distribution always remains a symmetric Gaussian, with its radius increasing with squeezing parameter $ r $. The reason for this behavior can be seen from the density matrix (\ref{photondens}) which shows that the state is a completely mixed state in the mode Fock states.  For $ r=0 $, the state is a vacuum state, and smoothly evolves towards a mixed state involving all mode Fock state numbers.  In contrast, the  dominant behavior of the atomic entangled state is to broaden along the equator, but not in the longitudinal direction.  This is as expected as in our expression (\ref{eq:aes}) we only perform
the first entangling step producing correlations in the $\phi $ direction but not $ \theta $.  A second measurement would produce the correlations in the $ \theta $ direction, with a corresponding broadening in this direction.  The spin and mode $Q$-functions show a general similarity (Figs. \ref{fig:q-css-marginal}(a)(b) and \ref{fig:q-optical-marginal}(a)(b)), particularly in the short time region where the HP approximation is valid.

\subsection{Ensemble-ensemble correlations}

One of the most striking features of the two-mode squeezed state is the presence of non-local correlations between the ensembles, make it useful for tasks such as quantum teleportation.  For an infinitely squeezed two-mode squeezed state (\ref{eprstate}), projecting on a $ | k \rangle $ or $ | k_x \rangle $ state on one mode collapses the other mode into the same state.  In this section we examine the correlations between the modes due to the entangling procedure, and compare them to a genuine two-mode squeezing operation. 

\begin{figure}
    \includegraphics[width=0.48\textwidth]{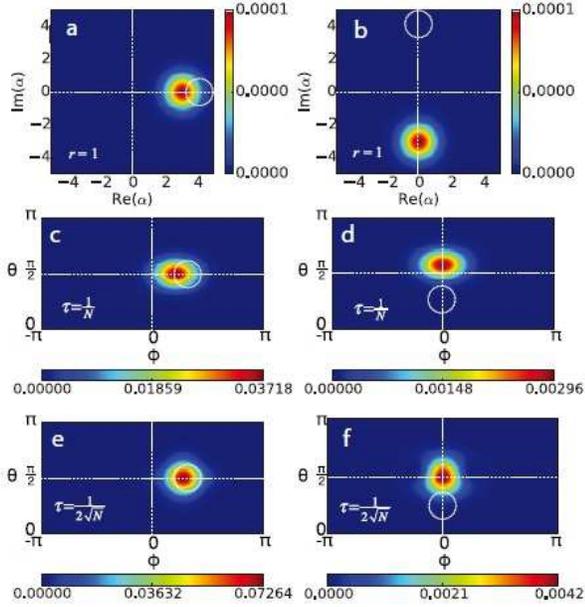}
    \caption{(a)(b) Joint $Q$-functions  (\ref{tmssqjoint}) for a genuine two-mode squeezed state with finite squeezing $r=1.0$. The second mode is projected on the coherence states centered at positions (a) $\alpha_2 = 4 $ and  (b) $ \alpha_2 = 4i $. (c)(d)(e)(f) Joint $Q$-distributions for the entangled atomic state Eq. (\ref{eq:q-joint-css}), for various point on the Bloch sphere as defined by the polar angle $\theta$ and the azimuthal angle $\phi$, and entanglement time $\tau$: 
		(c) $\tau = 1/N$, $\theta_2=\frac{\pi}{2}, \phi_2=1.0$, 
		(d) $\tau = 1/N$, $\theta_2=\frac{\pi}{4}, \phi_2=0$, 
		(e) $\tau = 1/2\sqrt{N}$, $\theta_2=\frac{\pi}{2}, \phi_2=1.0$, 
		(f) $\tau = 1/2\sqrt{N}$, $\theta_2=\frac{\pi}{4}, \phi_2=0 $. 
		The number of particles in each ensemble is $N=20$. The position of the projected state on the second ensemble/mode are shown with the white dotted circle.  }
    \label{fig:q-css-joint}
\end{figure}

First let us analyze what is expected for the $Q$-function of a genuine two-mode squeezed state.  Evaluating the $Q$-function (for bosonic modes) with respect to (\ref{tmsphoton}) gives
\begin{align}
Q(\alpha_1,\alpha_2)&  = \frac{1- \tanh^2 (r)}{\pi^2} \nonumber \\
& \times e^{ -  \tanh (r) | \alpha_1 - \alpha_2^*|^2 } e^{- (1-\tanh(r)) (|\alpha_1|^2+|\alpha_2|^2)} .
\label{tmssqjoint}
\end{align}
In the limit of strong squeezing $ r \rightarrow \infty $, we have $ \tanh (r) \rightarrow 1 $, and only the first exponential factor is present. If we now set $ \alpha_2 $ to a constant and consider the $Q$-distribution on mode 1, we observe that it is centered around $ \alpha_2^* $.  In terms of position and momentum, this corresponds to 
\begin{align}
x_1 = \text{Re}(\alpha_1) = \text{Re}(\alpha_2^*) = x_2 \nonumber \\
p_1 = \text{Im}(\alpha_1) = \text{Im}(\alpha_2^*) = -p_2 ,
\label{cvcorrelations}
\end{align}
i.e., the positions are correlated and momenta are anti-correlated. For finite squeezing, the additional Gaussian factor in (\ref{tmssqjoint}) suppresses the correlation, and the relations (\ref{cvcorrelations}) tend to saturate at finite values (see Fig. \ref{fig:q-css-joint}(a)(b)).

With this in mind, let us examine the $Q$-functions of the atomic entangled states.  
We first examine the spin $Q$-functions, calculating the overlap of the state with respect to coherent spin states (\ref{qfuncall}), for which we obtain
\begin{multline}
Q(\theta_1,\phi_1,\theta_2,\phi_2) =\\
= \frac{1}{\pi 4^N}\Big|\sum_{k_2,k_1}\cos\big(\frac{\theta_1}{2}\big)^{k_1}e^{i\phi_1(N-k_1)}\cos\big(\frac{\theta_2}{2}\big)^{k_2}e^{i\phi_2(N-k_2)}\\
\sin\big(\frac{\theta_1}{2}\big)^{N-k_1}\sin\big(\frac{\theta_2}{2}\big)^{N-k_2} C_N^{k_1}C_N^{k_2}A(k_1,k_2) \Big|^2.
\label{eq:q-joint-css}
\end{multline}
The a entangling times are shown in 
Fig. \ref{fig:q-css-joint}(c)(d)(e)(f), for various choices of interaction times $ \tau $ and coherent spin state positions $ \theta_2, \phi_2 $.  We choose the coherent state centers on ensemble 2 according to where the marginal distributions have a significant probability.  Choosing $ \theta_2, \phi_2 $ that are outside the marginal distribution corresponds to very low probability events, giving distributions that are more difficult to interpret.  In Fig. \ref{fig:q-css-joint}(c), we choose a coherent state center displaced along the equator in the positive direction.  We see that as expected the coherent spin state follows the chosen angle $ \phi_2 $ by the same amount.  This is the same behavior as would be expected of a two-mode squeezed state.  In Fig. \ref{fig:q-css-joint}(d), we vary the polar angle $ \theta_2 $ instead.  For small times $ \tau = 1/N $, we see that the distributions shows a weak anit-correlation in the angle.  
The anti-correlation is weaker than in the $ \phi $ direction, and does not preserve the symmetrical nature of the $Q$-function. This is again due to the fact that we only include the first measurement in our analysis, which produces the correlations in the $\phi $ direction.  It is however interesting that a despite not putting in the correlation at all, some anti-correlation is present.  For longer times $ \tau = 1/\sqrt{N} $, the $ \theta $ correlation becomes weaker, although the $\phi $ correlation remains in place.

To better quantify the correlation and anti-correlation between the two ensembles, we take the spin expectation values on ensemble 1 after projecting a coherent spin state on ensemble 2. Denoting the projector on the second ensemble as  $P(\theta_2,\phi_2) = | \theta_2, \phi_2 \rangle \rangle \langle \langle \theta_2, \phi_2 |$, the expectation values are calculated as
\begin{equation}
\langle J^{y,z}_1 (\theta_2,\phi_2) \rangle = \frac{\langle \Psi | J^{y,z}_1 P(\theta_2,\phi_2) | \Psi \rangle}{\langle \Psi |P(\theta_2,\phi_2) |  \Psi \rangle},
\end{equation}
where we require a normalization of the state as a projection is being made.  We obtain expressions for the average spins as
\begin{align}
\langle & J^z_1  (\theta_2,\phi_2) \rangle = \frac{1}{4^N {\cal N}_{\theta\phi }} \sum_{k_1,k_2,k'_2} C_N^{k_1} C_N^{k_2}C_N^{k_2'} (N-2k_1) \nonumber    \\
& \times  e^{i\phi (k_2' - k_2)} A(k_1,k_2) A(k_1,k_2') \cos^{k_2 + k_2'} (\tfrac{\theta}{2})
\sin^{2N - k_2 - k_2'} (\tfrac{\theta}{2}) ,
\end{align}
and
\begin{multline}
\langle J^y_1  (\theta_2,\phi_2) \rangle = \frac{1}{ {\cal N}_{\theta\phi }} \sum_{k_1,k_2,k'_2}C_N^{k_2}C_N^{k_2'} \sqrt{C_N^{k_1} } A(k_1,k_2)\\
\times \Big[i\sqrt{ (N-k_1)(k_1+1)}\sqrt{C_N^{k_1+1} } A(k_1+1,k_2')\\
- i\sqrt{  k_1(N-k_1+1)}\sqrt{C_N^{k_1-1} } A(k_1-1,k_2') \Big]\\
\times \cos^{k_2 + k_2'} (\tfrac{\theta}{2}) e^{i\phi (k_2' - k_2)}  \sin^{2N - k_2 - k_2'}  (\tfrac{\theta}{2}) ,
\end{multline}
where $ {\cal N}_{\theta\phi } $ is the normalization factor, common for both expectation values,
\begin{align}
  {\cal N}_{\theta\phi } & = \langle \Psi | P(\theta_2,\phi_2) | \Psi \rangle \nonumber \\
& = \frac{1}{2^N} \sum_{k_1,k_2,k'_2} C_N^{k_1} C_N^{k_2}C_N^{k_2'} A(k_1,k_2) A(k_1,k_2') \nonumber \\
&  \times \cos^{k_2 + k_2'}  (\tfrac{\theta}{2})  e^{i\phi (k_2' - k_2)}  \sin^{2N - k_2 - k_2'} (\tfrac{\theta}{2}).
\end{align}

\begin{figure}
    \includegraphics[width=0.48\textwidth]{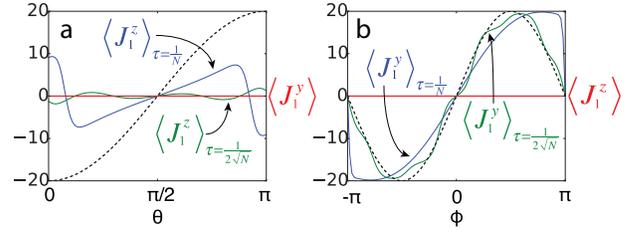}
    \caption{Expectation values $\langle J^z_1 \rangle $ and $\langle J^y_1 \rangle $  as functions of (a)  $\theta$, with fixed $\phi = 0$  and (b) $\phi$ with fixed $\theta = \pi/2$, for different entanglement times, $\tau=1/N$ and $\tau=1/2\sqrt{N}$. Dashed lines correspond to the expectation values using the HP approximation.
    $\langle J^y_1 \rangle$ as function of $\theta$ and  $\langle J^z_1 \rangle$ as function of $\phi$ are equal to zero for all entanglement times. Parameters used are $ N = 20 $ throughout.}
    \label{fig:exp-values}
\end{figure}

The expectation values are shown in Fig. \ref{fig:exp-values} for two different entanglement times, $\tau=\frac{1}{N}$ and $\tau = \frac{1}{2\sqrt{N}}$. 
In Fig. \ref{fig:exp-values}(b) we show the correlation with $ \phi_2 $.  As expected we see a positive correlation between the coherent spin state $ \phi_2 $ and the average spin $ \langle J^y_1 \rangle $.  According to (\ref{cvcorrelations}), for a genuine two-mode squeezed state we should observe under the HP approximation, Eq. (\ref{eq:cvmapping}),
\begin{align}
\langle J^y_1 \rangle & = \langle J^y_2 \rangle = N \sin \theta_2 \sin \phi_2 \nonumber \\
\langle J^z_1 \rangle & = -\langle J^z_2 \rangle = -N \cos \theta_2  ,
\label{idealtwomodecor}
\end{align}
where the expectation value involves the projection $P(\theta_2,\phi_2) $. For both the interaction times, the $ \langle J^y_1 \rangle $ expectations follow the expected sine curve.  The  $ \langle J^z_1 \rangle = 0$ indicating that the state remains on the equator of the Bloch sphere at all times.  The longer times start to deviate from the genuine two-mode squeezed state, due to the non-Gaussian characteristics that are developing in the state. For the $ \theta $ correlations (see Fig. \ref{fig:exp-values}(a)), we see the anti-correlations are generally weaker as expected.  While in the genuine CV regime one would not expect to see correlations at all (as only the first measurement is included), we nevertheless see a positive slope in which an increase in $\theta$ from the ground state $\theta = \frac{\pi}{2}$, corresponding to a negative change in the mapped momentum operator, $p $ results in a positive expectation value. The anti-correlations diminish with larger interaction times, hence is generally a weaker effect. Similarly to the $ \phi $ correlations, the $\langle J^y \rangle$ is equal to zero for all entanglement times, indicating there is no correlation with $ \phi $,
as expected.

\section{Entanglement}
\label{sec:entanglement}

In the previous section, we showed that as the interaction times grow longer, the states evolve into complex states with non-Gaussian properties.  For small interaction times the state has the characteristics of a two-mode squeezed state, but for longer times the state deviates from this and the expected correlations degrade.  One may be tempted to conclude from this observation that entanglement is degrading in the system. This is in fact untrue as the two-mode squeezing interaction is only an approximation at short times, and at longer times the entangling scheme produces a different type of entanglement at longer times. Due to the large number of degrees of freedom in an ensemble, it is possible to have many different types of entangled state, which are not necessarily simply described as a two-mode squeezed state. This is a similar situation to Ref. \cite{Byrnes2013}, where the system evolves according to a $ J^z_1 J^z_2 $ interaction, and a variety of different types of entangled states are produced. 

\begin{figure}
    \includegraphics[width=0.48\textwidth]{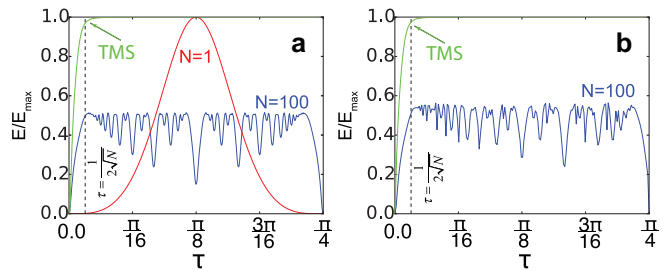}
    \caption{Entanglement as quantified by the von Neumann entropy of the atomic entangled state.  Measurement outcomes of the photon fields (a) $n_c = n_d = 50$, (b) $n_c=40$, $n_d=60$ are used when the particle number is $N=100$. The entanglement in the two-mode squeezed state (\ref{tmsspin}) labeled by TMS is shown for comparison. 
		The single qubit case has $n_c=n_d=1$. }
    \label{fig:entanglement}
\end{figure}

To show this explicitly, we plot the entanglement as quantified by the von Neumann entropy
\begin{align}
E = - \rho_1 \log_2 \rho_1 
\end{align}
where the density matrix is given in (\ref{partialtrace}).  The results are shown in Fig. \ref{fig:entanglement}.  We see that the entanglement shows a remarkably complex behavior with large dips at certain times. In Fig. \ref{fig:entanglement}(a) the measured photon numbers are equal, and the entanglement is symmetric around the characteristic time $\tau=\pi/8$. For $ N = 1$ (i.e. a single atom in each ensemble) we see that we recover maximal entanglement at $ \tau = \pi/8 $, where the state has the form of a Bell state.  The saturation of the entanglement at around half the maximal entanglement is exactly the same behavior as that seen in Ref. \cite{Byrnes2013}, which arose due to the binomial factors present in the wavefunction, which is also present in our case (\ref{eq:aes}).  The binomial factors create an uneven superposition of states which only involve spin Fock states that are centered around $ k_1, k_2 = N/2 $ with width $ \sim O(\sqrt{N}) $.  We attribute the saturation here to the same origin.  We also plot the entanglement of an ideal two-mode squeezed state (\ref{tmsspin}), using the mapping (\ref{squeezingmapping}) to convert the squeezing parameter $ r $ to interaction time.  For the ideal two-mode squeezed state, the entanglement increases towards a maximally entangled state, as can be directly seen from (\ref{tmsspin}) where there is an equal superposition of pairs of spin Fock states.  Such a state is never attained in our protocol due to the binomial factors as discussed above.

As the measured photon numbers are changed, we see in Fig. \ref{fig:entanglement}(b) that the entanglement follows overall the similar behavior, but different amplitudes of the dips in the entanglement. As the change in photon numbers does not change the arguments in Eq. (\ref{eq:a}), the dips appear at the same times. This behavior is remarkably similar of the fractal "devil's crevasse" entanglement produced by a $J^z_1 J^z_2$ interaction \cite{Byrnes2013}. Although in this case the interaction is not precisely the same, the Hamiltonian (\ref{eq:hamiltonian}) shares the same form and we attribute the similar entanglement to similar resonances of coherent state as seen in Ref. \cite{Byrnes2013}.  The complex structure of dips in the entanglement occur only after the characteristic time $ \tau = 1/2 \sqrt{N} $. Prior to this time there is a monotonic increase of the entanglement, with little variation on the photon number outcomes.  We thus expect that for times $ \tau > 1/2 \sqrt{N} $ there will be a strong dependence of the state on the obtained photon number.  Statistical averaging of states at this time will likely produce states with lower purity, which will degrade the entanglement, if no postselection is used.  In contrast, for earlier times  $ \tau < 1/2 \sqrt{N} $, there should be less dependence on the photon number.  In the limit of very short interaction times  $ \tau \sim 1/N $ the scheme reduces that to the CV approximated case, which should have very little photon number dependence to the entangled state that is created.  Such a behavior is consistent with the results found in Ref. \cite{Byrnes2013}, where the states where robust in the presence of decoherence for entangling times in the region  $ \tau \sim 1/N $, but sensitive for longer times  $ \tau \sim O(1) $.  The crossover between decoherence robust and sensitive regimes occurred at times $ \tau \sim 1/\sqrt{N} $, which is consistent with the results we observe here.

\section{Summary and conclusion}
\label{sec:discussion}

We have presented a model and analyzed the quantum state for the light-mediated entangling procedure as described in Refs. \cite{Julsgaard2001a,Kuzmich2000,Duan2001,Duan2000,N.J.CerfG.Leuchs} with the modification that we restrict us to modeling interaction in one direction.  The main benefit of our approach as compared to previous theoretical descriptions is that it does not make the HP approximation, where one of the spin degrees of freedom is treated as a c-number.  This allows us to investigate the beyond-CV regime, where the state starts to develop non-Gaussian characteristics.  We have investigated this state from the point of view of the Fock number probability distributions, the marginal and joint $Q$-function distributions, and entanglement between the ensembles.  From the simulations we find that the entangled state behaves like a two-mode squeezed state for short entanglement times of $\tau \sim 1/N$. Further evolving in time, new behavior is seen where the entangled coherent spin states become non-Gaussian in the marginal $Q$-function and wrap around the equator of the Bloch sphere. As the system enters the non-Gaussian regime, the correlations tend to degrade, where the simple CV correlations between coherent states breaks down.  Additionally, we have found an analytical relation between the squeezing parameter when using the HP approximation and the entanglement time of the atomic entangled state, which allows us to compare the genuine two-mode squeezed state to the atomic state. This gives a theoretical maximum squeezing that is possible, defined as where the HP approximation starts to break down. 

The diminishing of the two-mode squeezed correlations is not due to additional effects such as decoherence entering in the calculation -- our calculations do not take these into account and thus may be considered to be the ``ideal'' behavior.  This is evident by looking at the entanglement between the ensembles as shown in Fig. \ref{fig:entanglement}.  In fact, the regime where the HP approximation is valid $ \tau \sim 1/N $ in fact relatively has small amounts of entanglement compared to longer interaction times.  Thus entanglement is certainly present at all times, but it is not a two-mode squeezed state, but a more complex type.  Due to the large number of degrees of freedom of an ensemble, there are many different types of entanglement that are possible, in addition to a two-mode squeezed state. These have a non-Gaussian nature, as shown by the $Q$-functions, and potentially have large amounts of entanglement in them.  At particular times such as $ \tau = \pi/8 $ it is possible to write exactly as an entangled Schrodinger cat-Bell states (\ref{eq:catstate-odd}) and (\ref{eq:catstate-even}).  Such deviations from the HP approximation can also potentially occur in other experimental configurations than those considered here, for example those in Refs. \cite{peise16,peise16b} where two-mode squeezed states are induced by atomic interactions between hyperfine states.  The degree of deviation depends upon the degree of depopulation of the mode which is treated classically.

In this paper we did not include the effects of decoherence, as we found that the pure state properties are already rather rich, and have not been analyzed fully to our knowledge. In a realistic setting, the main sources of decoherence would be photon and atomic losses, which we expect to degrade the entanglement as observed in Ref. \cite{Byrnes2013}.  If the same behavior is observed as the $ J^z_1 J^z_2 $ interaction, then for times $ \tau > 1/\sqrt{N} $ decoherence effects are greatly enhanced due to the generation of Schrodinger cat-like states, which are known to be highly susceptible to decoherence. This scenario is quite likely due to the similar nature of the interaction Hamiltonian $ H \propto S^z (J^z_1 + J^z_2 ) $.  This suggests that there is a window of interaction times $ 1/N \lesssim \tau \lesssim 1/\sqrt{N} $ which is beyond the CV regime and shows non-Gaussian characteristics, yet is still stable under decoherence. One requirement would be that the entanglement state between the atoms would be relatively invariant to the photon counting outcomes $ n_c, n_d $.  To our knowledge this has yet to be observed experimentally, and would display rich physics with potential applications to quantum metrology and computing \cite{gerving2012,strobel2014,Byrnes2015,Byrnes2012}.

\section{acknowledgements}
O. P. and T. B. thank Qiongyi He and Yumang Jing for discussions. This work is supported by the Shanghai Research Challenge Fund, New York University Global Seed Grants for Collaborative Research, National Natural Science Foundation of China (Grant No. 61571301), the Thousand Talents Program for Distinguished Young Scholars (Grant No. D1210036A), and the NSFC Research Fund for International Young Scientists (Grant No. 11650110425). 
\appendix

\section{Derivation of two-mode squeezed probability distribution}

Here we explain how the probability distribution of (\ref{tmsspin}) can be evaluated to (\ref{tmss-final}).   In the region where the system is strongly polarized, near the extreme values $k=0,N$, the transformation elements $\langle k_x|k \rangle$ can be well approximated as harmonic oscillator functions, as described in Ref. \cite{Rowe2001},
\begin{align}
\langle k_{x}|k \rangle \approx \Psi_{m}(x) = (-1)^{m}  e^{-x^2/2} H_{m}(x) .
\end{align}
In this transformation, $m = \frac{N}{2}-|\frac{N}{2}-k^{x}|$, $x_{1,2}=-\frac{1}{\sqrt{2N}}(2k_{1,2}-N)$ and $H_{m}(x)$ is the $ m$th Hermite polynomial.
Armed with this we change basis from $k_x$ to $k$, and obtain,
\begin{align}
|\text{TMS (mode)}&\rangle  \propto \sum_{k_1 k_2 k_x}  \lambda^{k_x} \langle k_1| k_x\rangle \langle k_2 | k_x\rangle |k_1 \rangle | k_2\rangle \nonumber \\
 = & \sqrt{1-\lambda^2} \sum_{k_1 k_2 k_x} \lambda^{k_x}  \Psi_{m}(x_1) \Psi_{m}(x_2) |k_1, k_2\rangle
\end{align}
where $ \lambda = \tanh (r) $. Using the completeness relation for Hermite functions, the sum over $k_x$ can be expressed in terms of Mehler kernels \cite{Viskov2008}, which states that
\begin{multline}
\sum_{m=0}^{\infty} \lambda^{m} \Psi_m(x_1) \Psi_m(x_1) = \frac{1}{\sqrt{\pi(1-\lambda^2)}} \\
\times \exp\Big[-\frac{1-\lambda}{1+\lambda} \frac{(x_1+x_2)^2}{4} -\frac{1+\lambda}{1-\lambda} \frac{(x_1 - x_2)^2}{4} \Big].
\end{multline}
Using the above definitions for $x_1$ and $x_2$, and denoting $\alpha = \frac{1-\lambda}{1+\lambda} = e^{-2r}$, the two-mode squeezed state probability density in number representation, $p(k_1, k_2) = |\langle k_1 |\langle k_2 | \text{TMS (mode)} \rangle |^2$  can be expressed in terms of Gaussian functions,
\begin{multline}
p(k_1, k_2)  \propto \exp\Big[-\alpha \frac{(2k_1 + 2k_2 -2N)^2}{4N} -\frac{(2k_1 - 2k_2)^2}{4N\alpha} \Big],
\end{multline}
which after rearranging gives 
\begin{align}
p(k_1,k_2) \propto \exp\Big[-\frac{\alpha}{2N} \big((2k_{1} - N)^2 + (2k_{2} -N)^2\big) \nonumber \\
+\frac{1}{N}(\frac{1}{\alpha} -\alpha)(k_{1} + k_{2} -N)^2\Big].
\end{align}
Some elementary manipulations give (\ref{tmss-final}).  
%

\begin{thebibliography}{39}%
\makeatletter
\providecommand \@ifxundefined [1]{%
 \@ifx{#1\undefined}
}%
\providecommand \@ifnum [1]{%
 \ifnum #1\expandafter \@firstoftwo
 \else \expandafter \@secondoftwo
 \fi
}%
\providecommand \@ifx [1]{%
 \ifx #1\expandafter \@firstoftwo
 \else \expandafter \@secondoftwo
 \fi
}%
\providecommand \natexlab [1]{#1}%
\providecommand \enquote  [1]{``#1''}%
\providecommand \bibnamefont  [1]{#1}%
\providecommand \bibfnamefont [1]{#1}%
\providecommand \citenamefont [1]{#1}%
\providecommand \href@noop [0]{\@secondoftwo}%
\providecommand \href [0]{\begingroup \@sanitize@url \@href}%
\providecommand \@href[1]{\@@startlink{#1}\@@href}%
\providecommand \@@href[1]{\endgroup#1\@@endlink}%
\providecommand \@sanitize@url [0]{\catcode `\\12\catcode `\$12\catcode
  `\&12\catcode `\#12\catcode `\^12\catcode `\_12\catcode `\%12\relax}%
\providecommand \@@startlink[1]{}%
\providecommand \@@endlink[0]{}%
\providecommand \url  [0]{\begingroup\@sanitize@url \@url }%
\providecommand \@url [1]{\endgroup\@href {#1}{\urlprefix }}%
\providecommand \urlprefix  [0]{URL }%
\providecommand \Eprint [0]{\href }%
\providecommand \doibase [0]{http://dx.doi.org/}%
\providecommand \selectlanguage [0]{\@gobble}%
\providecommand \bibinfo  [0]{\@secondoftwo}%
\providecommand \bibfield  [0]{\@secondoftwo}%
\providecommand \translation [1]{[#1]}%
\providecommand \BibitemOpen [0]{}%
\providecommand \bibitemStop [0]{}%
\providecommand \bibitemNoStop [0]{.\EOS\space}%
\providecommand \EOS [0]{\spacefactor3000\relax}%
\providecommand \BibitemShut  [1]{\csname bibitem#1\endcsname}%
\let\auto@bib@innerbib\@empty
\bibitem [{\citenamefont {O'Connell}\ \emph {et~al.}(2010)\citenamefont
  {O'Connell}, \citenamefont {Hofheinz}, \citenamefont {Ansmann}, \citenamefont
  {Bialczak}, \citenamefont {Lenander}, \citenamefont {Lucero}, \citenamefont
  {Neeley}, \citenamefont {Sank}, \citenamefont {Wang}, \citenamefont {Weides},
  \citenamefont {Wenner}, \citenamefont {Martinis},\ and\ \citenamefont
  {Cleland}}]{OConnell2010}%
  \BibitemOpen
  \bibfield  {author} {\bibinfo {author} {\bibfnamefont {A.~D.}\ \bibnamefont
  {O'Connell}}, \bibinfo {author} {\bibfnamefont {M.}~\bibnamefont {Hofheinz}},
  \bibinfo {author} {\bibfnamefont {M.}~\bibnamefont {Ansmann}}, \bibinfo
  {author} {\bibfnamefont {R.~C.}\ \bibnamefont {Bialczak}}, \bibinfo {author}
  {\bibfnamefont {M.}~\bibnamefont {Lenander}}, \bibinfo {author}
  {\bibfnamefont {E.}~\bibnamefont {Lucero}}, \bibinfo {author} {\bibfnamefont
  {M.}~\bibnamefont {Neeley}}, \bibinfo {author} {\bibfnamefont
  {D.}~\bibnamefont {Sank}}, \bibinfo {author} {\bibfnamefont {H.}~\bibnamefont
  {Wang}}, \bibinfo {author} {\bibfnamefont {M.}~\bibnamefont {Weides}},
  \bibinfo {author} {\bibfnamefont {J.}~\bibnamefont {Wenner}}, \bibinfo
  {author} {\bibfnamefont {J.~M.}\ \bibnamefont {Martinis}}, \ and\ \bibinfo
  {author} {\bibfnamefont {A.~N.}\ \bibnamefont {Cleland}},\ }\href
  {http://dx.doi.org/10.1038/nature08967
  http://www.nature.com/nature/journal/v464/n7289/suppinfo/nature08967{\_}S1.html}
  {\bibfield  {journal} {\bibinfo  {journal} {Nature}\ }\textbf {\bibinfo
  {volume} {464}},\ \bibinfo {pages} {697} (\bibinfo {year}
  {2010})}\BibitemShut {NoStop}%
\bibitem [{\citenamefont {Julsgaard}\ \emph {et~al.}(2001)\citenamefont
  {Julsgaard}, \citenamefont {Kozhekin},\ and\ \citenamefont
  {Polzik}}]{Julsgaard2001a}%
  \BibitemOpen
  \bibfield  {author} {\bibinfo {author} {\bibfnamefont {B.}~\bibnamefont
  {Julsgaard}}, \bibinfo {author} {\bibfnamefont {A.}~\bibnamefont {Kozhekin}},
  \ and\ \bibinfo {author} {\bibfnamefont {E.~S.}\ \bibnamefont {Polzik}},\
  }\href {http://dx.doi.org/10.1038/35096524} {\bibfield  {journal} {\bibinfo
  {journal} {Nature}\ }\textbf {\bibinfo {volume} {413}},\ \bibinfo {pages}
  {400} (\bibinfo {year} {2001})}\BibitemShut {NoStop}%
\bibitem [{\citenamefont {Riedel}\ \emph {et~al.}(2010)\citenamefont {Riedel},
  \citenamefont {B{\"{o}}hi}, \citenamefont {Li}, \citenamefont {H{\"{a}}nsch},
  \citenamefont {Sinatra},\ and\ \citenamefont {Treutlein}}]{Riedel2010}%
  \BibitemOpen
  \bibfield  {author} {\bibinfo {author} {\bibfnamefont {M.~F.}\ \bibnamefont
  {Riedel}}, \bibinfo {author} {\bibfnamefont {P.}~\bibnamefont {B{\"{o}}hi}},
  \bibinfo {author} {\bibfnamefont {Y.}~\bibnamefont {Li}}, \bibinfo {author}
  {\bibfnamefont {T.~W.}\ \bibnamefont {H{\"{a}}nsch}}, \bibinfo {author}
  {\bibfnamefont {A.}~\bibnamefont {Sinatra}}, \ and\ \bibinfo {author}
  {\bibfnamefont {P.}~\bibnamefont {Treutlein}},\ }\href
  {http://dx.doi.org/10.1038/nature08988
  http://www.nature.com/nature/journal/v464/n7292/suppinfo/nature08988{\_}S1.html}
  {\bibfield  {journal} {\bibinfo  {journal} {Nature}\ }\textbf {\bibinfo
  {volume} {464}},\ \bibinfo {pages} {1170} (\bibinfo {year}
  {2010})}\BibitemShut {NoStop}%
\bibitem [{\citenamefont {Krauter}\ \emph {et~al.}(2013)\citenamefont
  {Krauter}, \citenamefont {Salart}, \citenamefont {Muschik}, \citenamefont
  {Petersen}, \citenamefont {Shen}, \citenamefont {Fernholz},\ and\
  \citenamefont {Polzik}}]{Krauter2013}%
  \BibitemOpen
  \bibfield  {author} {\bibinfo {author} {\bibfnamefont {H.}~\bibnamefont
  {Krauter}}, \bibinfo {author} {\bibfnamefont {D.}~\bibnamefont {Salart}},
  \bibinfo {author} {\bibfnamefont {C.~a.}\ \bibnamefont {Muschik}}, \bibinfo
  {author} {\bibfnamefont {J.~M.}\ \bibnamefont {Petersen}}, \bibinfo {author}
  {\bibfnamefont {H.}~\bibnamefont {Shen}}, \bibinfo {author} {\bibfnamefont
  {T.}~\bibnamefont {Fernholz}}, \ and\ \bibinfo {author} {\bibfnamefont
  {E.~S.}\ \bibnamefont {Polzik}},\ }\href {\doibase 10.1038/nphys2631}
  {\bibfield  {journal} {\bibinfo  {journal} {Nature Physics}\ }\textbf
  {\bibinfo {volume} {9}},\ \bibinfo {pages} {400} (\bibinfo {year}
  {2013})}\BibitemShut {NoStop}%
\bibitem [{\citenamefont {Pezz{\`e}}\ \emph {et~al.}(2016)\citenamefont
  {Pezz{\`e}}, \citenamefont {Smerzi}, \citenamefont {Oberthaler},
  \citenamefont {Schmied},\ and\ \citenamefont {Treutlein}}]{pezze16}%
  \BibitemOpen
  \bibfield  {author} {\bibinfo {author} {\bibfnamefont {L.}~\bibnamefont
  {Pezz{\`e}}}, \bibinfo {author} {\bibfnamefont {A.}~\bibnamefont {Smerzi}},
  \bibinfo {author} {\bibfnamefont {M.~K.}\ \bibnamefont {Oberthaler}},
  \bibinfo {author} {\bibfnamefont {R.}~\bibnamefont {Schmied}}, \ and\
  \bibinfo {author} {\bibfnamefont {P.}~\bibnamefont {Treutlein}},\ }\href@noop
  {} {\bibfield  {journal} {\bibinfo  {journal} {arxiv}\ ,\ \bibinfo {pages}
  {1609.01609}} (\bibinfo {year} {2016})}\BibitemShut {NoStop}%
\bibitem [{\citenamefont {Schlosshauer}(2005)}]{RevModPhys.76.1267}%
  \BibitemOpen
  \bibfield  {author} {\bibinfo {author} {\bibfnamefont {M.}~\bibnamefont
  {Schlosshauer}},\ }\href {\doibase 10.1103/RevModPhys.76.1267} {\bibfield
  {journal} {\bibinfo  {journal} {Rev. Mod. Phys.}\ }\textbf {\bibinfo {volume}
  {76}},\ \bibinfo {pages} {1267} (\bibinfo {year} {2005})}\BibitemShut
  {NoStop}%
\bibitem [{\citenamefont {Zurek}(2003)}]{RevModPhys.75.715}%
  \BibitemOpen
  \bibfield  {author} {\bibinfo {author} {\bibfnamefont {W.~H.}\ \bibnamefont
  {Zurek}},\ }\href {\doibase 10.1103/RevModPhys.75.715} {\bibfield  {journal}
  {\bibinfo  {journal} {Rev. Mod. Phys.}\ }\textbf {\bibinfo {volume} {75}},\
  \bibinfo {pages} {715} (\bibinfo {year} {2003})}\BibitemShut {NoStop}%
\bibitem [{\citenamefont {Byrnes}(2013)}]{Byrnes2013}%
  \BibitemOpen
  \bibfield  {author} {\bibinfo {author} {\bibfnamefont {T.}~\bibnamefont
  {Byrnes}},\ }\href {\doibase 10.1103/PhysRevA.88.023609} {\bibfield
  {journal} {\bibinfo  {journal} {Physical Review A}\ }\textbf {\bibinfo
  {volume} {88}},\ \bibinfo {pages} {023609} (\bibinfo {year}
  {2013})}\BibitemShut {NoStop}%
\bibitem [{\citenamefont {Byrnes}\ \emph {et~al.}(2015)\citenamefont {Byrnes},
  \citenamefont {Rosseau}, \citenamefont {Khosla}, \citenamefont {Pyrkov},
  \citenamefont {Thomasen}, \citenamefont {Mukai}, \citenamefont {Koyama},
  \citenamefont {Abdelrahman},\ and\ \citenamefont {Ilo-Okeke}}]{Byrnes2015}%
  \BibitemOpen
  \bibfield  {author} {\bibinfo {author} {\bibfnamefont {T.}~\bibnamefont
  {Byrnes}}, \bibinfo {author} {\bibfnamefont {D.}~\bibnamefont {Rosseau}},
  \bibinfo {author} {\bibfnamefont {M.}~\bibnamefont {Khosla}}, \bibinfo
  {author} {\bibfnamefont {A.}~\bibnamefont {Pyrkov}}, \bibinfo {author}
  {\bibfnamefont {A.}~\bibnamefont {Thomasen}}, \bibinfo {author}
  {\bibfnamefont {T.}~\bibnamefont {Mukai}}, \bibinfo {author} {\bibfnamefont
  {S.}~\bibnamefont {Koyama}}, \bibinfo {author} {\bibfnamefont
  {A.}~\bibnamefont {Abdelrahman}}, \ and\ \bibinfo {author} {\bibfnamefont
  {E.}~\bibnamefont {Ilo-Okeke}},\ }\href {\doibase
  10.1016/j.optcom.2014.08.017} {\bibfield  {journal} {\bibinfo  {journal}
  {Optics Communications}\ }\textbf {\bibinfo {volume} {337}},\ \bibinfo
  {pages} {102} (\bibinfo {year} {2015})}\BibitemShut {NoStop}%
\bibitem [{\citenamefont {Byrnes}\ \emph {et~al.}(2012)\citenamefont {Byrnes},
  \citenamefont {Wen},\ and\ \citenamefont {Yamamoto}}]{Byrnes2012}%
  \BibitemOpen
  \bibfield  {author} {\bibinfo {author} {\bibfnamefont {T.}~\bibnamefont
  {Byrnes}}, \bibinfo {author} {\bibfnamefont {K.}~\bibnamefont {Wen}}, \ and\
  \bibinfo {author} {\bibfnamefont {Y.}~\bibnamefont {Yamamoto}},\ }\href
  {\doibase 10.1103/PhysRevA.85.040306} {\bibfield  {journal} {\bibinfo
  {journal} {Physical Review A}\ }\textbf {\bibinfo {volume} {85}},\ \bibinfo
  {pages} {040306(R)} (\bibinfo {year} {2012})}\BibitemShut {NoStop}%
\bibitem [{\citenamefont {Filip}(2013)}]{filip13}%
  \BibitemOpen
  \bibfield  {author} {\bibinfo {author} {\bibfnamefont {R.}~\bibnamefont
  {Filip}},\ }\href {\doibase 10.1103/PhysRevA.87.042308} {\bibfield  {journal}
  {\bibinfo  {journal} {Phys. Rev. A}\ }\textbf {\bibinfo {volume} {87}},\
  \bibinfo {pages} {042308} (\bibinfo {year} {2013})}\BibitemShut {NoStop}%
\bibitem [{\citenamefont {Ilo-Okeke}\ and\ \citenamefont
  {Byrnes}(2014)}]{Ilo-Okeke2014}%
  \BibitemOpen
  \bibfield  {author} {\bibinfo {author} {\bibfnamefont {E.~O.}\ \bibnamefont
  {Ilo-Okeke}}\ and\ \bibinfo {author} {\bibfnamefont {T.}~\bibnamefont
  {Byrnes}},\ }\href {\doibase 10.1103/PhysRevLett.112.233602} {\bibfield
  {journal} {\bibinfo  {journal} {Physical Review Letters}\ }\textbf {\bibinfo
  {volume} {112}},\ \bibinfo {pages} {233602} (\bibinfo {year}
  {2014})}\BibitemShut {NoStop}%
\bibitem [{\citenamefont {Pyrkov}\ and\ \citenamefont
  {Byrnes}(2014{\natexlab{a}})}]{1367-2630-16-7-073038}%
  \BibitemOpen
  \bibfield  {author} {\bibinfo {author} {\bibfnamefont {A.~N.}\ \bibnamefont
  {Pyrkov}}\ and\ \bibinfo {author} {\bibfnamefont {T.}~\bibnamefont
  {Byrnes}},\ }\href {http://stacks.iop.org/1367-2630/16/i=7/a=073038}
  {\bibfield  {journal} {\bibinfo  {journal} {New Journal of Physics}\ }\textbf
  {\bibinfo {volume} {16}},\ \bibinfo {pages} {73038} (\bibinfo {year}
  {2014}{\natexlab{a}})}\BibitemShut {NoStop}%
\bibitem [{\citenamefont {Tichy}\ \emph {et~al.}(2016)\citenamefont {Tichy},
  \citenamefont {Park}, \citenamefont {Kang}, \citenamefont {Jeong},\ and\
  \citenamefont {M\o{}lmer}}]{tichy16}%
  \BibitemOpen
  \bibfield  {author} {\bibinfo {author} {\bibfnamefont {M.~C.}\ \bibnamefont
  {Tichy}}, \bibinfo {author} {\bibfnamefont {C.-Y.}\ \bibnamefont {Park}},
  \bibinfo {author} {\bibfnamefont {M.}~\bibnamefont {Kang}}, \bibinfo {author}
  {\bibfnamefont {H.}~\bibnamefont {Jeong}}, \ and\ \bibinfo {author}
  {\bibfnamefont {K.}~\bibnamefont {M\o{}lmer}},\ }\href {\doibase
  10.1103/PhysRevA.93.042314} {\bibfield  {journal} {\bibinfo  {journal} {Phys.
  Rev. A}\ }\textbf {\bibinfo {volume} {93}},\ \bibinfo {pages} {042314}
  (\bibinfo {year} {2016})}\BibitemShut {NoStop}%
\bibitem [{\citenamefont {Pyrkov}\ and\ \citenamefont
  {Byrnes}(2014{\natexlab{b}})}]{PhysRevA.90.062336}%
  \BibitemOpen
  \bibfield  {author} {\bibinfo {author} {\bibfnamefont {A.~N.}\ \bibnamefont
  {Pyrkov}}\ and\ \bibinfo {author} {\bibfnamefont {T.}~\bibnamefont
  {Byrnes}},\ }\href {\doibase 10.1103/PhysRevA.90.062336} {\bibfield
  {journal} {\bibinfo  {journal} {Phys. Rev. A}\ }\textbf {\bibinfo {volume}
  {90}},\ \bibinfo {pages} {62336} (\bibinfo {year}
  {2014}{\natexlab{b}})}\BibitemShut {NoStop}%
\bibitem [{\citenamefont {Furusawa}\ \emph {et~al.}(1998)\citenamefont
  {Furusawa}, \citenamefont {S{\o}rensen}, \citenamefont {Braunstein},
  \citenamefont {Fuchs}, \citenamefont {Kimble},\ and\ \citenamefont
  {Polzik}}]{Furusawa23101998}%
  \BibitemOpen
  \bibfield  {author} {\bibinfo {author} {\bibfnamefont {A.}~\bibnamefont
  {Furusawa}}, \bibinfo {author} {\bibfnamefont {J.~L.}\ \bibnamefont
  {S{\o}rensen}}, \bibinfo {author} {\bibfnamefont {S.~L.}\ \bibnamefont
  {Braunstein}}, \bibinfo {author} {\bibfnamefont {C.~A.}\ \bibnamefont
  {Fuchs}}, \bibinfo {author} {\bibfnamefont {H.~J.}\ \bibnamefont {Kimble}}, \
  and\ \bibinfo {author} {\bibfnamefont {E.~S.}\ \bibnamefont {Polzik}},\
  }\href {\doibase 10.1126/science.282.5389.706} {\bibfield  {journal}
  {\bibinfo  {journal} {Science}\ }\textbf {\bibinfo {volume} {282}},\ \bibinfo
  {pages} {706} (\bibinfo {year} {1998})}\BibitemShut {NoStop}%
\bibitem [{\citenamefont {Braunstein}\ and\ \citenamefont {van
  Loock}(2005)}]{Braunstein2005}%
  \BibitemOpen
  \bibfield  {author} {\bibinfo {author} {\bibfnamefont {S.~L.}\ \bibnamefont
  {Braunstein}}\ and\ \bibinfo {author} {\bibfnamefont {P.}~\bibnamefont {van
  Loock}},\ }\href {\doibase 10.1103/RevModPhys.77.513} {\bibfield  {journal}
  {\bibinfo  {journal} {Reviews of Modern Physics}\ }\textbf {\bibinfo {volume}
  {77}},\ \bibinfo {pages} {513} (\bibinfo {year} {2005})}\BibitemShut
  {NoStop}%
\bibitem [{\citenamefont {Kuzmich}\ and\ \citenamefont
  {Polzik}(2000)}]{Kuzmich2000}%
  \BibitemOpen
  \bibfield  {author} {\bibinfo {author} {\bibfnamefont {A.}~\bibnamefont
  {Kuzmich}}\ and\ \bibinfo {author} {\bibfnamefont {E.~S.}\ \bibnamefont
  {Polzik}},\ }\href {\doibase 10.1103/PhysRevLett.85.5639} {\bibfield
  {journal} {\bibinfo  {journal} {Physical review letters}\ }\textbf {\bibinfo
  {volume} {85}},\ \bibinfo {pages} {5639} (\bibinfo {year}
  {2000})}\BibitemShut {NoStop}%
\bibitem [{\citenamefont {Duan}\ \emph {et~al.}(2001)\citenamefont {Duan},
  \citenamefont {Lukin}, \citenamefont {Cirac},\ and\ \citenamefont
  {Zoller}}]{Duan2001}%
  \BibitemOpen
  \bibfield  {author} {\bibinfo {author} {\bibfnamefont {L.~M.}\ \bibnamefont
  {Duan}}, \bibinfo {author} {\bibfnamefont {M.~D.}\ \bibnamefont {Lukin}},
  \bibinfo {author} {\bibfnamefont {J.~I.}\ \bibnamefont {Cirac}}, \ and\
  \bibinfo {author} {\bibfnamefont {P.}~\bibnamefont {Zoller}},\ }\href
  {\doibase 10.1038/35106500} {\bibfield  {journal} {\bibinfo  {journal}
  {Nature}\ }\textbf {\bibinfo {volume} {414}},\ \bibinfo {pages} {413}
  (\bibinfo {year} {2001})}\BibitemShut {NoStop}%
\bibitem [{\citenamefont {Duan}\ \emph {et~al.}(2000)\citenamefont {Duan},
  \citenamefont {Cirac}, \citenamefont {Zoller},\ and\ \citenamefont
  {Polzik}}]{Duan2000}%
  \BibitemOpen
  \bibfield  {author} {\bibinfo {author} {\bibfnamefont {L.~M.}\ \bibnamefont
  {Duan}}, \bibinfo {author} {\bibfnamefont {J.~I.}\ \bibnamefont {Cirac}},
  \bibinfo {author} {\bibfnamefont {P.}~\bibnamefont {Zoller}}, \ and\ \bibinfo
  {author} {\bibfnamefont {E.~S.}\ \bibnamefont {Polzik}},\ }\href {\doibase
  10.1103/PhysRevLett.85.5643} {\bibfield  {journal} {\bibinfo  {journal}
  {Phys. Rev. Lett.}\ }\textbf {\bibinfo {volume} {85}},\ \bibinfo {pages}
  {5643} (\bibinfo {year} {2000})}\BibitemShut {NoStop}%
\bibitem [{\citenamefont {Cerf}\ \emph {et~al.}(2007)\citenamefont {Cerf},
  \citenamefont {Leuchs},\ and\ \citenamefont {Polzik}}]{N.J.CerfG.Leuchs}%
  \BibitemOpen
  \bibfield  {author} {\bibinfo {author} {\bibfnamefont {N.}~\bibnamefont
  {Cerf}}, \bibinfo {author} {\bibfnamefont {G.}~\bibnamefont {Leuchs}}, \ and\
  \bibinfo {author} {\bibfnamefont {E.}~\bibnamefont {Polzik}},\ }\href
  {\doibase 10.1142/9781860948169} {\emph {\bibinfo {title} {{Quantum
  information with continuous variables of atoms and light}}}},\ Vol.~\bibinfo
  {volume} {1}\ (\bibinfo  {publisher} {Wiley},\ \bibinfo {year} {2007})\ pp.\
  \bibinfo {pages} {1829--1841}\BibitemShut {NoStop}%
\bibitem [{\citenamefont {Gerving}\ \emph {et~al.}(2012)\citenamefont
  {Gerving}, \citenamefont {Hoang}, \citenamefont {Land}, \citenamefont
  {Anquez}, \citenamefont {Hamley},\ and\ \citenamefont
  {Chapman}}]{gerving2012}%
  \BibitemOpen
  \bibfield  {author} {\bibinfo {author} {\bibfnamefont {C.~S.}\ \bibnamefont
  {Gerving}}, \bibinfo {author} {\bibfnamefont {T.~M.}\ \bibnamefont {Hoang}},
  \bibinfo {author} {\bibfnamefont {B.~J.}\ \bibnamefont {Land}}, \bibinfo
  {author} {\bibfnamefont {M.}~\bibnamefont {Anquez}}, \bibinfo {author}
  {\bibfnamefont {C.~D.}\ \bibnamefont {Hamley}}, \ and\ \bibinfo {author}
  {\bibfnamefont {M.~S.}\ \bibnamefont {Chapman}},\ }\href {\doibase
  10.1038/ncomms2179} {\bibfield  {journal} {\bibinfo  {journal} {Nature
  Comm.}\ }\textbf {\bibinfo {volume} {3}},\ \bibinfo {pages} {1169} (\bibinfo
  {year} {2012})}\BibitemShut {NoStop}%
\bibitem [{\citenamefont {Strobel}\ \emph {et~al.}(2014)\citenamefont
  {Strobel}, \citenamefont {Muessel}, \citenamefont {Linnemann}, \citenamefont
  {Zibold}, \citenamefont {Hume}, \citenamefont {Pezze}, \citenamefont
  {Smerzi},\ and\ \citenamefont {Oberthaler}}]{strobel2014}%
  \BibitemOpen
  \bibfield  {author} {\bibinfo {author} {\bibfnamefont {H.}~\bibnamefont
  {Strobel}}, \bibinfo {author} {\bibfnamefont {W.}~\bibnamefont {Muessel}},
  \bibinfo {author} {\bibfnamefont {D.}~\bibnamefont {Linnemann}}, \bibinfo
  {author} {\bibfnamefont {T.}~\bibnamefont {Zibold}}, \bibinfo {author}
  {\bibfnamefont {D.~B.}\ \bibnamefont {Hume}}, \bibinfo {author}
  {\bibfnamefont {L.}~\bibnamefont {Pezze}}, \bibinfo {author} {\bibfnamefont
  {a.}~\bibnamefont {Smerzi}}, \ and\ \bibinfo {author} {\bibfnamefont {M.~K.}\
  \bibnamefont {Oberthaler}},\ }\href {\doibase 10.1126/science.1250147}
  {\bibfield  {journal} {\bibinfo  {journal} {Science}\ }\textbf {\bibinfo
  {volume} {345}},\ \bibinfo {pages} {424} (\bibinfo {year}
  {2014})}\BibitemShut {NoStop}%
\bibitem [{\citenamefont {Barontini}\ \emph {et~al.}(2015)\citenamefont
  {Barontini}, \citenamefont {Hohmann}, \citenamefont {Haas}, \citenamefont
  {Est{\`e}ve},\ and\ \citenamefont {Reichel}}]{barontini15}%
  \BibitemOpen
  \bibfield  {author} {\bibinfo {author} {\bibfnamefont {G.}~\bibnamefont
  {Barontini}}, \bibinfo {author} {\bibfnamefont {L.}~\bibnamefont {Hohmann}},
  \bibinfo {author} {\bibfnamefont {F.}~\bibnamefont {Haas}}, \bibinfo {author}
  {\bibfnamefont {J.}~\bibnamefont {Est{\`e}ve}}, \ and\ \bibinfo {author}
  {\bibfnamefont {J.}~\bibnamefont {Reichel}},\ }\href {\doibase
  10.1126/science.aaa0754} {\bibfield  {journal} {\bibinfo  {journal}
  {Science}\ }\textbf {\bibinfo {volume} {349}},\ \bibinfo {pages} {1317}
  (\bibinfo {year} {2015})}\BibitemShut {NoStop}%
\bibitem [{\citenamefont {McConnell}\ \emph {et~al.}(2015)\citenamefont
  {McConnell}, \citenamefont {Zhang}, \citenamefont {Hu}, \citenamefont
  {{\'C}uk},\ and\ \citenamefont {Vuleti{\'c}}}]{mcconnell15}%
  \BibitemOpen
  \bibfield  {author} {\bibinfo {author} {\bibfnamefont {R.}~\bibnamefont
  {McConnell}}, \bibinfo {author} {\bibfnamefont {H.}~\bibnamefont {Zhang}},
  \bibinfo {author} {\bibfnamefont {J.}~\bibnamefont {Hu}}, \bibinfo {author}
  {\bibfnamefont {S.}~\bibnamefont {{\'C}uk}}, \ and\ \bibinfo {author}
  {\bibfnamefont {V.}~\bibnamefont {Vuleti{\'c}}},\ }\href@noop {} {\bibfield
  {journal} {\bibinfo  {journal} {Nature}\ }\textbf {\bibinfo {volume} {519}},\
  \bibinfo {pages} {439} (\bibinfo {year} {2015})}\BibitemShut {NoStop}%
\bibitem [{\citenamefont {Bohnet}\ \emph {et~al.}(2016)\citenamefont {Bohnet},
  \citenamefont {Sawyer}, \citenamefont {Britton}, \citenamefont {Wall},
  \citenamefont {Rey}, \citenamefont {Foss-Feig},\ and\ \citenamefont
  {Bollinger}}]{bohnet16}%
  \BibitemOpen
  \bibfield  {author} {\bibinfo {author} {\bibfnamefont {J.~G.}\ \bibnamefont
  {Bohnet}}, \bibinfo {author} {\bibfnamefont {B.~C.}\ \bibnamefont {Sawyer}},
  \bibinfo {author} {\bibfnamefont {J.~W.}\ \bibnamefont {Britton}}, \bibinfo
  {author} {\bibfnamefont {M.~L.}\ \bibnamefont {Wall}}, \bibinfo {author}
  {\bibfnamefont {A.~M.}\ \bibnamefont {Rey}}, \bibinfo {author} {\bibfnamefont
  {M.}~\bibnamefont {Foss-Feig}}, \ and\ \bibinfo {author} {\bibfnamefont
  {J.~J.}\ \bibnamefont {Bollinger}},\ }\href {\doibase
  10.1126/science.aad9958} {\bibfield  {journal} {\bibinfo  {journal}
  {Science}\ }\textbf {\bibinfo {volume} {352}},\ \bibinfo {pages} {1297}
  (\bibinfo {year} {2016})}\BibitemShut {NoStop}%
\bibitem [{\citenamefont {Husimi}(1940)}]{citeulike:7123199}%
  \BibitemOpen
  \bibfield  {author} {\bibinfo {author} {\bibfnamefont {K.}~\bibnamefont
  {Husimi}},\ }\href
  {http://www.journalarchive.jst.go.jp/japanese/jnlabstract{\_}ja.php?cdjournal=ppmsj1919{\&}{\#}38;cdvol=22{\&}{\#}38;noissue=4{\&}{\#}38;startpage=264}
  {\bibfield  {journal} {\bibinfo  {journal} {Nippon Sugaku-Buturigakkwai Kizi
  Dai 3 Ki}\ }\textbf {\bibinfo {volume} {22}},\ \bibinfo {pages} {264}
  (\bibinfo {year} {1940})}\BibitemShut {NoStop}%
\bibitem [{\citenamefont {Gross}(2012)}]{Gross2012}%
  \BibitemOpen
  \bibfield  {author} {\bibinfo {author} {\bibfnamefont {C.}~\bibnamefont
  {Gross}},\ }\href {\doibase 10.1088/0953-4075/45/10/103001} {\bibfield
  {journal} {\bibinfo  {journal} {Journal of Physics B: Atomic, Molecular and
  Optical Physics}\ }\textbf {\bibinfo {volume} {45}},\ \bibinfo {pages}
  {103001} (\bibinfo {year} {2012})}\BibitemShut {NoStop}%
\bibitem [{\citenamefont {Hammerer}\ \emph {et~al.}(2010)\citenamefont
  {Hammerer}, \citenamefont {S{\o}rensen},\ and\ \citenamefont
  {Polzik}}]{Hammerer2010}%
  \BibitemOpen
  \bibfield  {author} {\bibinfo {author} {\bibfnamefont {K.}~\bibnamefont
  {Hammerer}}, \bibinfo {author} {\bibfnamefont {A.~S.}\ \bibnamefont
  {S{\o}rensen}}, \ and\ \bibinfo {author} {\bibfnamefont {E.~S.}\ \bibnamefont
  {Polzik}},\ }\href {\doibase 10.1103/RevModPhys.82.1041} {\bibfield
  {journal} {\bibinfo  {journal} {Reviews of Modern Physics}\ }\textbf
  {\bibinfo {volume} {82}},\ \bibinfo {pages} {1041} (\bibinfo {year}
  {2010})}\BibitemShut {NoStop}%
\bibitem [{\citenamefont {Tinkham}(2003)}]{tinkham2003group}%
  \BibitemOpen
  \bibfield  {author} {\bibinfo {author} {\bibfnamefont {M.}~\bibnamefont
  {Tinkham}},\ }\href {https://books.google.com.hk/books?id=r4GIU2wJCAEC}
  {\emph {\bibinfo {title} {{Group Theory and Quantum Mechanics}}}},\ Dover
  Books on Chemistry and Earth Sciences\ (\bibinfo  {publisher} {Dover
  Publications},\ \bibinfo {year} {2003})\BibitemShut {NoStop}%
\bibitem [{\citenamefont {Thompson}(2008)}]{thompson2008angular}%
  \BibitemOpen
  \bibfield  {author} {\bibinfo {author} {\bibfnamefont {W.~J.}\ \bibnamefont
  {Thompson}},\ }\href {https://books.google.com.hk/books?id=0NMjkQnQN6oC}
  {\emph {\bibinfo {title} {{Angular Momentum}}}}\ (\bibinfo  {publisher}
  {Wiley},\ \bibinfo {year} {2008})\BibitemShut {NoStop}%
\bibitem [{\citenamefont {Kuzmich}\ \emph {et~al.}(2007)\citenamefont
  {Kuzmich}, \citenamefont {Bigelow},\ and\ \citenamefont
  {Mandel}}]{Kuzmich2007}%
  \BibitemOpen
  \bibfield  {author} {\bibinfo {author} {\bibfnamefont {A.}~\bibnamefont
  {Kuzmich}}, \bibinfo {author} {\bibfnamefont {N.~P.}\ \bibnamefont
  {Bigelow}}, \ and\ \bibinfo {author} {\bibfnamefont {L.}~\bibnamefont
  {Mandel}},\ }\href {\doibase 10.1209/epl/i1998-00277-9} {\bibfield  {journal}
  {\bibinfo  {journal} {Europhys. Lett.}\ }\textbf {\bibinfo {volume} {42}},\
  \bibinfo {pages} {481} (\bibinfo {year} {2007})}\BibitemShut {NoStop}%
\bibitem [{\citenamefont {Takahashi}\ \emph {et~al.}(1999)\citenamefont
  {Takahashi}, \citenamefont {Honda}, \citenamefont {Tanaka}, \citenamefont
  {Toyoda}, \citenamefont {Ishikawa},\ and\ \citenamefont
  {Yabuzaki}}]{Takahashi1999}%
  \BibitemOpen
  \bibfield  {author} {\bibinfo {author} {\bibfnamefont {Y.}~\bibnamefont
  {Takahashi}}, \bibinfo {author} {\bibfnamefont {K.}~\bibnamefont {Honda}},
  \bibinfo {author} {\bibfnamefont {N.}~\bibnamefont {Tanaka}}, \bibinfo
  {author} {\bibfnamefont {K.}~\bibnamefont {Toyoda}}, \bibinfo {author}
  {\bibfnamefont {K.}~\bibnamefont {Ishikawa}}, \ and\ \bibinfo {author}
  {\bibfnamefont {T.}~\bibnamefont {Yabuzaki}},\ }\href {\doibase
  10.1103/PhysRevA.60.4974} {\bibfield  {journal} {\bibinfo  {journal}
  {Physical Review A}\ }\textbf {\bibinfo {volume} {60}},\ \bibinfo {pages}
  {4974} (\bibinfo {year} {1999})}\BibitemShut {NoStop}%
\bibitem [{\citenamefont {Takeuchi}\ \emph {et~al.}(2005)\citenamefont
  {Takeuchi}, \citenamefont {Ichihara}, \citenamefont {Takano}, \citenamefont
  {Kumakura}, \citenamefont {Yabuzaki},\ and\ \citenamefont
  {Takahashi}}]{Takeuchi2005}%
  \BibitemOpen
  \bibfield  {author} {\bibinfo {author} {\bibfnamefont {M.}~\bibnamefont
  {Takeuchi}}, \bibinfo {author} {\bibfnamefont {S.}~\bibnamefont {Ichihara}},
  \bibinfo {author} {\bibfnamefont {T.}~\bibnamefont {Takano}}, \bibinfo
  {author} {\bibfnamefont {M.}~\bibnamefont {Kumakura}}, \bibinfo {author}
  {\bibfnamefont {T.}~\bibnamefont {Yabuzaki}}, \ and\ \bibinfo {author}
  {\bibfnamefont {Y.}~\bibnamefont {Takahashi}},\ }\href {\doibase
  10.1103/PhysRevLett.94.023003} {\bibfield  {journal} {\bibinfo  {journal}
  {Physical Review Letters}\ }\textbf {\bibinfo {volume} {94}},\ \bibinfo
  {pages} {023003} (\bibinfo {year} {2005})}\BibitemShut {NoStop}%
\bibitem [{\citenamefont {Ilo-Okeke}\ and\ \citenamefont
  {Byrnes}(2016)}]{Ilo-okeke2015}%
  \BibitemOpen
  \bibfield  {author} {\bibinfo {author} {\bibfnamefont {E.~O.}\ \bibnamefont
  {Ilo-Okeke}}\ and\ \bibinfo {author} {\bibfnamefont {T.}~\bibnamefont
  {Byrnes}},\ }\href@noop {} {\bibfield  {journal} {\bibinfo  {journal} {Phys.
  Rev. A}\ }\textbf {\bibinfo {volume} {94}},\ \bibinfo {pages} {013617}
  (\bibinfo {year} {2016})}\BibitemShut {NoStop}%
\bibitem [{\citenamefont {Peise}\ \emph {et~al.}(2016)\citenamefont {Peise},
  \citenamefont {Kruse}, \citenamefont {Lange}, \citenamefont {Lücke},
  \citenamefont {Pezze}, \citenamefont {Arlt}, \citenamefont {Ertmer},
  \citenamefont {Hammerer}, \citenamefont {Santos}, \citenamefont {Smerzi},\
  and\ \citenamefont {Klempt}}]{peise16}%
  \BibitemOpen
  \bibfield  {author} {\bibinfo {author} {\bibfnamefont {J.}~\bibnamefont
  {Peise}}, \bibinfo {author} {\bibfnamefont {I.}~\bibnamefont {Kruse}},
  \bibinfo {author} {\bibfnamefont {K.}~\bibnamefont {Lange}}, \bibinfo
  {author} {\bibfnamefont {B.}~\bibnamefont {Lücke}}, \bibinfo {author}
  {\bibfnamefont {L.}~\bibnamefont {Pezze}}, \bibinfo {author} {\bibfnamefont
  {J.}~\bibnamefont {Arlt}}, \bibinfo {author} {\bibfnamefont {W.}~\bibnamefont
  {Ertmer}}, \bibinfo {author} {\bibfnamefont {K.}~\bibnamefont {Hammerer}},
  \bibinfo {author} {\bibfnamefont {L.}~\bibnamefont {Santos}}, \bibinfo
  {author} {\bibfnamefont {A.}~\bibnamefont {Smerzi}}, \ and\ \bibinfo {author}
  {\bibfnamefont {C.}~\bibnamefont {Klempt}},\ }\href@noop {} {\bibfield
  {journal} {\bibinfo  {journal} {Nature Communications}\ }\textbf {\bibinfo
  {volume} {6}},\ \bibinfo {pages} {8984} (\bibinfo {year} {2016})}\BibitemShut
  {NoStop}%
\bibitem [{\citenamefont {Peise}\ \emph {et~al.}(2015)\citenamefont {Peise},
  \citenamefont {Lucke}, \citenamefont {Pezze}, \citenamefont {Deuretzbacher},
  \citenamefont {Ertmer}, \citenamefont {Arlt}, \citenamefont {Smerzi},
  \citenamefont {Santos},\ and\ \citenamefont {Klempt}}]{peise16b}%
  \BibitemOpen
  \bibfield  {author} {\bibinfo {author} {\bibfnamefont {J.}~\bibnamefont
  {Peise}}, \bibinfo {author} {\bibfnamefont {B.}~\bibnamefont {Lucke}},
  \bibinfo {author} {\bibfnamefont {L.}~\bibnamefont {Pezze}}, \bibinfo
  {author} {\bibfnamefont {F.}~\bibnamefont {Deuretzbacher}}, \bibinfo {author}
  {\bibfnamefont {W.}~\bibnamefont {Ertmer}}, \bibinfo {author} {\bibfnamefont
  {J.}~\bibnamefont {Arlt}}, \bibinfo {author} {\bibfnamefont {A.}~\bibnamefont
  {Smerzi}}, \bibinfo {author} {\bibfnamefont {L.}~\bibnamefont {Santos}}, \
  and\ \bibinfo {author} {\bibfnamefont {C.}~\bibnamefont {Klempt}},\
  }\href@noop {} {\bibfield  {journal} {\bibinfo  {journal} {Nature
  Communications}\ }\textbf {\bibinfo {volume} {6}},\ \bibinfo {pages} {6811}
  (\bibinfo {year} {2015})}\BibitemShut {NoStop}%
\bibitem [{\citenamefont {Rowe}\ \emph {et~al.}(2001)\citenamefont {Rowe},
  \citenamefont {Guise},\ and\ \citenamefont {Sanders}}]{Rowe2001}%
  \BibitemOpen
  \bibfield  {author} {\bibinfo {author} {\bibfnamefont {D.~J.}\ \bibnamefont
  {Rowe}}, \bibinfo {author} {\bibfnamefont {H.~D.}\ \bibnamefont {Guise}}, \
  and\ \bibinfo {author} {\bibfnamefont {B.~C.}\ \bibnamefont {Sanders}},\
  }\href@noop {} {\bibfield  {journal} {\bibinfo  {journal} {Journal of
  Mathematical Physics}\ }\textbf {\bibinfo {volume} {42}},\ \bibinfo {pages}
  {2315} (\bibinfo {year} {2001})}\BibitemShut {NoStop}%
\bibitem [{\citenamefont {Viskov}(2008)}]{Viskov2008}%
  \BibitemOpen
  \bibfield  {author} {\bibinfo {author} {\bibfnamefont {O.~V.}\ \bibnamefont
  {Viskov}},\ }\href {\doibase 10.1134/S1064562408010018} {\bibfield  {journal}
  {\bibinfo  {journal} {Doklady Mathematics}\ }\textbf {\bibinfo {volume}
  {77}},\ \bibinfo {pages} {1} (\bibinfo {year} {2008})}\BibitemShut {NoStop}%
\end{thebibliography}
%

%
%
\end{document}